\RequirePackage{ifpdf}
\documentclass[11pt,a4paper]{article}
\usepackage{amsmath,amssymb,epsfig,jheppubme, bm}
\usepackage{multirow,longtable,enumerate,mathtools}
\usepackage{pdflscape, array}
\usepackage[raggedrightboxes]{ragged2e}
\usepackage{changepage}
\usepackage{graphicx}
\usepackage[x11names]{xcolor}

\newtheorem{prop}{Proposition}
\newtheorem{conj}{Conjecture}

\font\cmss=cmss12

\newcommand\half{\frac12}

\newcommand\bi{\begin{itemize}}
\newcommand\ei{\end{itemize}}

\newcommand\bea{\begin{eqnarray}}
\newcommand\eea{\end{eqnarray}}
\newcommand\be{\begin{equation}}
\newcommand\ee{\end{equation}}

\newcommand\cT{{\cal T}}

\newcommand\cW{{\cal W}}
\newcommand\cV{{\cal V}}

\newcommand\cN{{\cal N}}

\newcommand\cD{{\cal D}}

\newcommand{\cO}{{\cal O}}

\newcommand{\cE}{{\cal E}}

\newcommand\ZZ{\hbox{Z\kern-.4emZ}}
\newcommand\sZZ{\hbox{\sevenfont Z\kern-.4emZ}}

\newcommand{\cM}{{\cal M}}

\newcommand{\eref}[1]{Eq.\,(\ref{#1})}
\newcommand{\Comment}[1]{{}}

\newcommand\Tstrut{\rule{0pt}{3.6ex}}

\def\IB{\relax{\rm I\kern-.18em B}}
\def\IC{{\relax\hbox{\kern.3em{\cmss I}$\kern-.4em{\rm C}$}}}
\def\ID{\relax{\rm I\kern-.18em D}}
\def\IE{\relax{\rm I\kern-.18em E}}
\def\IF{\relax{\rm I\kern-.18em F}}
\def\II{\relax{\rm I\kern-.18em I}}

\def\Id{\relax{1\kern-.32em 1}}
\def\IG{\relax\hbox{$\inbar\kern-.3em{\rm G}$}}
\def\IR{\relax{\rm I\kern-.18em R}}

\newcommand{\la}{\lambda}

\newcommand{\cI}{{\cal I}}
\newcommand{\cC}{{\cal C}}
\newcommand{\cZ}{{\cal Z}}

\newcommand{\mbf}[1]{\mathbf{#1}}
\newcommand{\mf}[1]{\mathfrak{#1}}

\title{Generalised 4d Partition Functions and Modular Differential Equations} 
\author[a]{A. Ramesh Chandra,}
\author[\,b]{Sunil Mukhi\,\footnote{Adjunct Professor, ICTS Bengaluru.}} 
\author[c,d]{and Palash Singh}
\affiliation[a]{International Centre for Theoretical Sciences,\\ Tata Institute of Fundamental Research,\\ Shivakote, Bengaluru 560089, India}
\affiliation[b]{Indian Institute of Science Education and Research,\\ Homi Bhabha Rd, Pashan, Pune 411 008, India} 
\affiliation[c]{Dipartimento di Fisica, Università di Milano--Bicocca,\\ Piazza della Scienza 3, I-20126 Milano, Italy}
\affiliation[d]{INFN, Sezione di Milano--Bicocca,\\ Piazza della Scienza 3, I-20126 Milano, Italy\\}

\emailAdd{ramesh.chandra@icts.res.in}
\emailAdd{sunil.mukhi@gmail.com}
\emailAdd{palash.singh@unimib.it}

\abstract{
We prove the equivalence of a class of generalised Schur partition functions $\cZ_G(q;\alpha)$ of 4d $\cN=2$ superconformal gauge theories to contour integral representations of vector-valued modular forms of the type that arise in 2d rational conformal field theories (RCFT). Concretely, we consider the $USp(2N)$ theory with $2N+2$ fundamental hypermultiplets and analytically prove that $\cZ_{USp(2N)}(q;\alpha)$ satisfies an order-$(N+1)$ modular linear differential equation (MLDE) with vanishing Wronskian index, explaining how the parameter $\alpha$ of the former determines the parameters of the latter. Several connections are made to characters of RCFTs including unitary ones. We then propose a two-parameter extension $\cZ_{USp(2N)}(q;\alpha,\beta)$ of the generalised Schur partition function. Finally, we relate the $\alpha=-k$ specialisation to quantum monodromy traces ${\rm Tr}\,M^k$ and formulate a conjecture linking their $k$-dependence to MLDEs.
}

\begin{document}

\maketitle

\parindent=0pt
\advance\parskip by 3pt

\section{Introduction and Summary}

The study of four-dimensional $\cN=2$ superconformal field theories (SCFT) has revealed a remarkably rich algebraic structure in their protected observables. The superconformal index is one such protected observable that plays a distinguished role in organising protected spectra, testing non-trivial dualities, and revealing rigid algebraic structures \cite{Kinney:2005ej,Romelsberger:2005eg}.

A particularly striking manifestation of such a rigid structure is the 4d/2d correspondence, which associates a vertex operator algebra (VOA) to any 4d $\cN=2$ SCFT via a cohomological construction \cite{Beem:2013sza}. For a unitary 4d $\cN=2$ SCFT, the associated VOA is necessarily non-unitary and encodes the spectrum and OPE data of a protected subsector of quarter-BPS operators, known as the Schur operators. A key consequence of this correspondence is that the Schur limit of the superconformal index \cite{Gadde:2011ik,Gadde:2011uv}, referred to as the {\it Schur index}, coincides with the vacuum character of the associated VOA. Moreover, in cases where the associated VOA is quasi-lisse \cite{Arakawa:2016hkg}, the Schur indices satisfy a modular linear differential equation (MLDE) \cite{Beem:2017ooy}.

Rather surprisingly, the Schur index of a 4d $\cN=2$ SCFT can also be computed by an infrared formula as the trace of the inverse of the Kontsevich-Soibelmann monodromy operator \cite{Buican:2015ina,Cordova:2015nma} (also see \cite{Cecotti:2010fi,Iqbal:2012xm,Buican:2015hsa,Cordova:2017mhb,Cordova:2017ohl,Buican:2017uka}). Within this framework, it has been observed in examples that traces of higher powers, say $k$, of the monodromy operator yield $q$-series that can be identified with vacuum characters of certain vertex algebras $\cV^{(k)}$ \cite{Cecotti:2015lab,Kim:2024dxu}. While $k=-1$ recovers the non-unitary VOA arising in the 4d/2d correspondence, the monodromy traces for $k>0$ encountered in many examples match the characters of unitary VOAs, including rational and logarithmic 2d conformal field theories.

Independent of these considerations, one can define a one-parameter generalisation of the Schur index of an $\cN=2$ superconformal gauge theory  $\cZ_G(q;\alpha)$ that was termed the {\it generalised Schur partition function} \cite{Deb:2025ypl}. These are defined via a double scaling limit of the full superconformal index in such a way that $\alpha=1$ trivially returns the usual Schur index of the SCFT. Based on empirical observations, it was suggested that the generalised Schur partition functions of certain gauge theories numerically coincide with Schur indices of different SCFTs, with the same rank, for finely tuned positive rational values of $\alpha$.

Our goal in the present work is to take a concrete step towards systematising some of these observations by mapping a family of generalised Schur partition functions to known contour integral representations of vector-valued modular forms (VVMF), possibly of logarithmic or quasi-modular type, that solve an MLDE \cite{Mathur:1988gt,Mukhi:1989qk,Mukhi:2019cpu}. Along the way, we analytically prove many of the numerically observed statements in \cite{Deb:2025ypl}.

Concretely, we consider the analytically continued generalised Schur partition function of the $USp(2N)$ SCFT with $2N+2$ fundamental hypermultiplets. 

Focusing on the $N=1$ case first, we prove that the $\cZ_{SU(2)}(q;\alpha)$ -- associated to $SU(2)$ with $4$ fundamental hypermultiplets -- satisfies a second order MLDE with vanishing Wronskian index for general $\alpha$. We formulate this precisely in Proposition \ref{prop:SU2} and provide a proof by mapping the standard gauge integral representation of $\cZ_{SU(2)}(q;\alpha)$ to a known contour integral representation that solves the corresponding MLDE: the MMS equation \cite{Mathur:1988rx,Mathur:1988na}. We supplement this by a discussion of many interesting examples, effectively proving the numerical observations made in \cite{Deb:2025ypl}, while also discovering several cases of a unitary RCFT character appearing at special values of $\alpha$.

Subsequently, we formulate an analogous proposition for general $N$, positing in Proposition \ref{prop:USp2N} that $\cZ_{USp(2N)}(q;\alpha)$ solves a one-parameter sub-family of the order-($N+1$) MLDE with vanishing Wronskian index. After providing a proof separately for the $N=2$ and the general $N$ case, we discuss several interesting families, in part proving the observations of \cite{Deb:2025ypl} for this case, and then finding some unitary RCFT characters for special values of $\alpha$. This analysis naturally leads to a conjecture for the two-parameter generalised Schur partition function that appears natural from the MLDE perspective. We then formulate Proposition \ref{prop:USp2N_2par}, provide a proof, and then discuss some special cases, in part showing that the complete classification of unitary three-character RCFTs of vanishing Wronskian index \cite{Das:2022uoe} can be obtained from the two-parameter extension of the generalised Schur partition function for $USp(4)$.

In Subsection \ref{subsec:monodromy}, we comment on the relation between the generalised Schur partition function and ${\rm Tr}\,
M^k$, where $M$ is the monodromy operator, for $\alpha=-k$. In the case of $USp(2N)$ SCFT with $N_f=2N+2$, we demonstrate that the central charges of the vertex algebras $\cV^{(k)}$ associated to ${\rm Tr}\,M^k$ \cite{Cecotti:2015lab,Kim:2024dxu} follow directly from our Proposition \ref{prop:USp2N}. Motivated by these observations, we formulate the general Conjecture \ref{conj1} that relates the $k$-dependent quantum monodromy traces to MLDEs.

Next, we discuss our proposal for the two-parameter extension of the generalised Schur partition function and leave its relation to the full superconformal index as an open problem. We conclude with remarks for general gauge groups $G$, where we expect analogous propositions to hold, and outline some of the technical challenges involved. The two trailing appendices collect many of the technical details that we use in our proof.
\newline

{\it Authors' note: This work appears concurrently with an independent study on related topics} \cite{Deb:2025ddc}{\it. The two papers were developed separately, though their release is coordinated.}

\section{Background review}
\label{sec:review}

In this section, we provide a basic review of the various four-dimensional and two-dimensional constructions of interest. We begin by a lightning review of the full superconformal index to set up notation. We then restrict to the Schur limit of the superconformal index and discuss the case of $\cN=2$ superconformal gauge theories. We conclude the 4d review by discussing the construction of the generalised Schur partition function. Switching gears, we first discuss the modular linear differential equation approach to RCFT characters, followed by a review of the contour integral representation of these characters in certain cases and their connection to MLDE.

\subsection{Superconformal Index}
\label{subsec:superconformal_index}

The general superconformal index of a four-dimensional $\cN=2$ superconformal field theory is a supersymmetric partition function that counts $1/8$ BPS cohomologies \cite{Kinney:2005ej,Romelsberger:2005eg} (also see \cite{Dolan:2008qi}). This depends on three fugacities coupled to three different combinations of charges in the superconformal algebra that commute with a chosen supercharge and its conformal superpartner, and thus survive in the BPS cohomology.

Concretely, the superconformal index is the $S^3\times S^1$ supersymmetric partition function
\begin{equation}\label{eq:superconfIndex_full}
    \cI(p,q,t,\mbf a) = {\rm Tr}_{S^3}\,(-1)^F\,p^{j_2-j_1-r} \,q^{j_1+j_2-r} \,t^{R+r} \prod_{i=1}^{{\rm rk}\,G_F} a_i^{\lambda_i} ~,
\end{equation}
where $j_1,j_2$ are the Cartan generators of $SU(2)\times SU(2)$ associated to the $S^3$, $R$ is the Cartan generator of the $SU(2)_R$ $R$-symmetry, $r$ is the charge of the $U(1)_r$ $R$-symmetry, and $\lambda_i$ are the Cartan generators of the $G_F$ flavour symmetry. We will only be interested in superconformal indices with no fugacity for the flavour symmetries, so we set all $a_i=1$.

The basic building blocks of 4d $\cN=2$ Lagrangian SCFTs are the $\cN=2$ hypermultiplets and vector multiplets. A general $\cN=2$ superconformal gauge theory with gauge group $G$ contains $\mathcal N=2$ vector multiplets in the adjoint representation of $G$ and $N_l$ hypermultiplets that transform in representations $\mathcal R_l$ of $G$. The contributions of these building blocks to the superconformal index are simply encoded in the single-letter indices \cite{Gadde:2011uv}
\begin{equation}\label{eq:singleletter_general}
    \begin{split}
        i_V(p,q,t,\mbf z) &= \left(-\frac{p}{1-p} - \frac{q}{1-q} + \frac{pq/t-t}{(1-p)(1-q)}\right)\chi_{\rm adj}(\mbf z) ~, \\[2mm]
        i_H^{\mathcal R}(p,q,t,\mbf z) &= \frac{\sqrt t-pq/\sqrt t}{(1-p)(1-q)} \Big(\chi_{\mathcal R}(\mbf z)+\chi_{\overline{\mathcal R}}(\mbf z)\Big) ~.
    \end{split}
\end{equation}

The full (unflavoured) superconformal index of the $G$ gauge theory with $N_l$ hypermultiplets in representations $R_l$ of $G$ can be expressed as
\begin{equation}\label{eq:full_superconfIndex_Lag}
    \cI(p,q,t) = \oint [{\rm d}\mbf z]_G \,{\rm PE}\left[i_V(p,q,t,\mbf  z)+\sum_l N_l\,i_H^{\mathcal R_l}(p,q,t,\mbf z)\right] ~,
\end{equation}
where $\oint[{\rm d}\mbf z]_G$ denotes the integral over the Haar measure of $G$ that projects onto gauge invariants of $G$. The symbol ${\rm PE}[\,\cdot\,]$ stands for Plethystic exponential defined as
\begin{equation}
    {\rm PE} \big[ f(x_1,x_2,\dots) \big] = {\rm Exp} \left( \sum_{n=1}^\infty \frac1n\,f(x_1^n,x_2^n,\dots) \right) ~,
\end{equation}
which generates all symmetric products of the function $f(x_1,x_2,\dots)$.

\subsection{Schur index}
\label{subsec:Schur_index}

There are many limits of this superconformal index that counts BPS cohomologies that preserve higher amounts of supersymmetry \cite{Gadde:2011uv}. One such limit is the Schur limit of the superconformal index, henceforth referred to as the {\it Schur index}, that receives contributions from states annihilated by more than one supercharge. More precisely, the Schur limit of the superconformal index counts quarter-BPS cohomologies and is defined as
\begin{equation}
    t = q \;,\;\text{ arbitrary $p$.}
\end{equation}
It can be easily checked that the charge associated to the fugacity $p$ also commutes with the supercharges that annihilate Schur operators, and hence the dependence on $p$ drops out. Therefore, the Schur index can be expressed as the following trace over the $S^3$ Hilbert space
\begin{equation}
    \widehat{\cI}(q) = {\rm Tr}_{S^3}\,(-1)^Fq^{\Delta-R} ~,
\end{equation}
where $\Delta$ is the scaling dimension.

\subsubsection{Relation to the 4d/2d correspondence and MLDE}
\label{subsubsec:Schur_4d2d_MLDE}

The Schur limit of the superconformal index is special, in part, because of the role it plays in the 4d/2d correspondence \cite{Beem:2013sza}. As the name suggests, the index receives contributions from the so-called Schur operators, that satisfy the constraint
\begin{equation}
    \Delta - (j_1 + j_2) - 2R = 0 \;,\; r + j_1 - j_2 = 0 ~.
\end{equation}
These are precisely the operators whose cohomology classes live in the vertex operator algebra associated to any given 4d $\cN=2$ SCFT in the context of the 4d/2d correspondence.\footnote{Note that this cohomology is distinct from the one used to define the superconformal index, and is specified by a particular choice of ``$\mathcal Q+\mathcal S$" nilpotent supercharge.} Under this correspondence, the (normalised) Schur index is identified with the vacuum character\,\footnote{Throughout this paper, we switch back and forth between referring to $\chi_0$ as the vacuum or the identity character. The latter terminology is appropriate for characters corresponding to unitary 2d RCFTs, as there is the smallest critical exponent is related to the central charge. For generic, possibly non-unitary, VOAs, there is no preferred ordering of the critical exponents and tht term 'vacuum' is more appropriate.}, $\chi_0(q)$, of the associated VOA\footnote{In a slight abuse of notation, we use the same $\cI(\,\cdot\,)$ to denote both the full superconformal index and the Schur index. The difference should be clear from the argument.}
\begin{equation}
    \mathcal I(q) :=q^{\frac{c_{\rm 4d}}{2}}\,\widehat{\mathcal I}(q) = \chi_0(q) = {\rm Tr}_{\rm VOA}\,q^{L_0-\frac{c_{2d}}{24}} ~,
\end{equation}
where we have introduced a normalisation factor in the normalised Schur index that depends purely on the $4d$ central charge. While it is quite natural to introduce this overall factor in the Schur index from the perspective of the 4d/2d correspondence, it was independently observed that it plays an important role in the modular properties of the Schur index \cite{Razamat:2012uv}.

An essential feature of this correspondence is the relation between the four-dimensional and two-dimensional central charge:
\begin{equation}
    c_{\rm 2d} = -12\,c_{\rm 4d} ~,
\end{equation}
which implies that the VOA associated to a unitary 4d $\cN=2$ SCFT is necessarily non-unitary. Furthermore, the Higgs branch of the 4d SCFT is conjecturally identified with the associated variety of the corresponding VOA. As shown in \cite{Beem:2017ooy}, this conjcture relies on the existence of a null vector in the VOA that, in some cases, was shown to give rise to a (possibly twisted) modular linear differential equation satisfied by the Schur index. More generally, it has been proven that vacuum characters of quasi-lisse VOA\footnote{A vertex operator algebra is said to be {\it quasi-lisse} if the associated variety has finitely many symplectic leaves. This condition is generally expected to hold for VOAs associated to 4d $\mathcal N=2$ SCFTs. \cite{Arakawa:2016hkg}} always satisfy a monic (Wronskian index, $\ell=0$) MLDE. While \cite{Arakawa:2016hkg} establishes the existence of such a monic MLDE, in practice the Schur index often satisfies a lower-order minimal $\ell\neq0$ MLDE, as was demonstrated in the context of 4d $\cN=2$ theories of class $\mathcal S$ in \cite{Beem:2021zvt}.

\subsubsection{Schur indices for gauge theories}
\label{subsubsec:Schur_Lagrangian}

Let us now consider a Lagrangian $\mathcal N=2$ SCFT with gauge algebra $G$ and $N_l$ hypermultiplets in representation $\mathcal R_l$ of $G$. The single-letter indices for the vector multiplet and hypermultiplets simplify in this limit and become
\begin{equation}
    i_V(q,\mbf z) = -\frac{2q}{1-q}\,\chi_{\rm adj}(\mbf z) \;,\; i_H^{\mathcal R}(q,\mbf z) = \frac{\sqrt q}{1-q} \left(\chi_{\mathcal R}(\mbf z)+\chi_{\overline{\mathcal R}}(\mbf z)\right) ~.
\end{equation}
Therefore, the Schur index can once again be expressed as an integral over the Haar measure of the Plethystic exponential of the single letter indices by specialising \eref{eq:full_superconfIndex_Lag},
\begin{equation}\label{eq:Schur_gauge_single}
    \cI(q) = q^{\frac{c_{\rm 4d}}{2}}\oint\big[{\rm d}\mbf z\big]_G\,{\rm PE}\left[-\frac{2q}{1-q}\chi_{\rm adj}(\mbf z) + \sum_lN_l\,\frac{\sqrt q}{1-q} \left(\chi_{\mathcal R_l}(\mbf z)+\chi_{\overline{\mathcal R}_l}(\mbf z)\right)\right] ~,
\end{equation}
where the integral is over the maximal torus of $G$ with the standard Haar measure
\begin{equation}
    \big[{\rm d}\mbf z\big]_G = \frac{1}{|W|}\prod_{j=1}^{{\rm rk}\,G}\frac{{\rm d}z_j}{2\pi i z_j} \,\Delta(\mbf z) \;,\; \text{ where } \Delta(\mbf z) \coloneq {\rm PE}\big[ {\rm rk}\,G-\chi_{\rm adj}(\mbf z) \big] ~,
\end{equation}
with $|W|$ being the dimension of the Weyl group of $G$. Note that in this case, the $c_{\rm 4d}$ central charge is given by
\begin{equation}
    c_{\rm 4d} = \frac{2\,n_V+n_H}{12} \;,\ \text{ where } n_V={\rm dim}\,\mf g, \,n_H=\sum_l({\rm dim}\,\mathcal R_l)\,N_l ~.
\end{equation}

Through standard identities relating $q$-Pochhammers and Jacobi theta functions reviewed in Appendix \ref{app:special_functions}, the Schur index of a Lagrangian $\cN=2$ SCFT in \eref{eq:full_superconfIndex_Lag} can be re-expressed as the following integral (see \cite{Pan:2021mrw}):
\begin{equation}\label{eq:Lag_SchurIndex}
    \begin{split}
        \cI(q) = \frac{(-i)^{{\rm rk}\,G-{\rm dim}\,G}\,\eta(\tau)^{3\,{\rm rk}\,G-{\rm dim}\,G}}{|W|} \oint\prod_{j=1}^{{\rm rk}\,\mathfrak g}\frac{{\rm d}z_j}{2\pi iz_j} \prod_{\bm\alpha\neq0} \vartheta_1(\bm\alpha\cdot\mathbf u,\tau) \prod_l\prod_{{\bm\rho}\in\mathcal R_l} \left(\frac{\eta(\tau)}{\vartheta_4( \bm\rho\cdot\mathbf u),\tau)}\right)^{N_l} ~,
    \end{split}
\end{equation}
where $\bm\alpha$ denotes the non-zero roots, $\bm\rho$ denotes the weights of the representation $\mathcal R_l$, and we define
\begin{equation}
    z_j \coloneq e^{2\pi i\,u_j} \;,\quad\forall\;j ~.
\end{equation}
Schematically, the contribution of the vector multiplet and the Haar measure in \eref{eq:full_superconfIndex_Lag} combine to produce the overall power of the Dedekind eta function $\eta(\tau)$ and the product over $\vartheta_1(\bm\alpha\cdot\mbf u,\tau)$ in \eref{eq:Lag_SchurIndex}. On the other hand, the hypermultiplet contribution can be simplifed into the ratio of $\eta(\tau)$ and $\vartheta_4(\bm\rho\cdot\mbf u,\tau)$. 

\subsection*{Example: $SU(N)$ with $N_f=2N$ fundamental hypermultiplets}

Let us consider a concrete example to demonstrate the expression \eref{eq:Lag_SchurIndex} for the Schur index. The $SU(N)$ SCFT with $2N$ fundamental hypermultiplets is the prototypical example of a 4d $\mathcal N=2$ superconformal gauge theory. The relevant group theoretic data for $SU(N)$ is
\begin{equation}
    \begin{split}
        {\rm rk}\,SU(N) = N-1 \;&,\; {\rm dim}\, SU(N) = N^2-1 \;,\; |W| = N! \;, \\
        \chi_{\rm fund}(\mbf z) = \sum_{i=1}^{N} z_i \;&,\; \chi_{\rm adj}(\mbf z) = N-1+\sum_{i\neq j}\frac{z_i}{z_j} \;,\; \text{where } \prod_{i=1}^N z_i = 1 ~.
    \end{split}
\end{equation}
We further note that the non-zero roots $\alpha_{ij}$, with $i\neq j$, act on the Cartan variables as $\alpha_{ij}(\mbf u)=u_i-u_j$. Substituting this data in \eref{eq:Lag_SchurIndex} and simplifying, one obtains
\begin{equation}
    \cI(q) = \frac{(-i)^{N(1-N)}}{N!}\,\eta(\tau)^{N^2+3N-2} \oint \prod_{j=1}^{N-1}\frac{{\rm d}z_j}{2\pi iz_j} \prod_{k\neq l}\vartheta_1(u_k-u_l,\tau)\prod_{k=1}^N \frac{1}{\vartheta_4^{2N}(u_k,\tau)} ~,
\end{equation}
where the fugacities are related by $\sum_{i=1}^N u_i=0$. It can be easily checked that the overall power of $q$ is
\begin{equation}
    q^{\frac{2N^2-1}{12}} = q^{\frac{c_{\rm 4d}}{2}} \;, \text{ where } c_{\rm 4d} = \frac{2(N^2-1)+2N^2}{12} ~.
\end{equation}

\subsection{Generalised Schur partition function}
\label{subsec:generalised_schur}

The generalised Schur partition function is defined as the following double scaling limit of the full superconformal index:
\begin{equation}\label{eq:gen_Schur_limit}
    p = 1 - \epsilon \;,\; t = (q\,p)^{1+\frac{\alpha}{{\rm log}\,q}\epsilon} \;,\; \text{ with } \epsilon\rightarrow0 ~.
\end{equation}
This can be best understood  as interpolating between two particular limits of the full index with an order of limits issue.

The first limit is setting $t\rightarrow q$ first and then taking $p \rightarrow1$. This is consistent with the Schur limit which leaves $p$ arbitrary. The second limit corresponds to taking $p\rightarrow t/q$ first and then taking the Schur limit $t\rightarrow q$. The limit $p \rightarrow t/q $ physically corresponds to giving a non-zero mass to all hypermultiplets and turning on a vacuum expectation value for all operators that parametrise the Coulomb branch of the moduli space of vacua of the 4d SCFT. The crucial observation in \cite{Deb:2025ypl} was that there is an order of limits issue here, and they defined \eref{eq:gen_Schur_limit} to interpolate between the former ($\alpha=1$) and the latter ($\alpha=0$).

The single letter indices of the vector multiplet and hypermultiplets take the following form in this limit:
\begin{equation}\label{eq:singleletter_genSchur}
    \begin{split}
        i_V(q,\mbf z;\alpha) &= \left(1-\alpha-\frac{2q}{1-q}\right)\,\chi_{\rm adj}(\mbf z) ~, \\
        i_H^{\mathcal R}(q,\mbf z;\alpha) &= \alpha\,\frac{\sqrt q}{1-q}\big(\chi_{\mathcal R}(\mbf z) + \chi_{\overline{\mathcal R}}(\mbf z)\big) ~.
    \end{split}
\end{equation}
The (un-normalised) generalised Schur partition function was then defined in \cite{Deb:2025ypl} to be
\begin{equation}\label{eq:NoNorm_GenSchur_def}
    \begin{split}
        \widehat{\cZ}_G(q;\alpha) &= \frac{1}{N(\alpha)\,|W|}\oint \prod_{j=1}^{{\rm rk}\,G}\frac{{\rm d}z_j}{2\pi i z_j}\,\Delta(\mbf z)\,{\rm PE}\big[i_V(q,\mbf z;\alpha) + \sum_l N_l\, i_H^{\mathcal R_l}(q,\mbf z;\alpha)\big] \\
        &= \frac{(q;q)^{2\,{\rm rk}\,G}}{N(\alpha)\,|W|}\oint \prod_{j=1}^{{\rm rk}\,G}\frac{{\rm d}z_j}{2\pi i z_j}\,\Delta(\mbf z)^\alpha \left(\frac{\prod_{\bm\alpha\neq0}(q\,e^{2\pi i\,\bm\alpha\cdot\mbf u};q)^2}{\prod_l\prod_{{\bm\rho}\in\mathcal R_l}(e^{2\pi i\,\bm\rho\cdot\mbf u}\sqrt q;q)^{N_l}}\right)^\alpha ~,
    \end{split}
\end{equation}
where recall that $z_j=e^{2\pi i\,u_j}$, while $N(\alpha)$ is a constant prefactor introduced by hand such that the leading order coefficient of this function is 1, and is defined to be\footnote{This is a trivial generalisation of the normalisation considered in \cite{Deb:2025ypl} to a general gauge group $G$.}
\begin{equation}
    N(\alpha) = \frac{1}{|W|}\oint \prod_{j=1}^{{\rm rk}\,G}\frac{{\rm d}z_j}{2\pi i z_j}\,\Delta(\mbf z)^{\alpha} ~.
\end{equation}
The second line in \eref{eq:NoNorm_GenSchur_def} can be obtained by evaluating the Plethystic exponential and is expressed in the same form as in \cite{Deb:2025ypl}. Some care is required as there is a divergence in the full superconformal index in the limit where $p\,q=t$ that sneaks into this calculation \cite{Gaiotto:2012xa}. The appropriate regularisation furnishes the factor $(q;q)^{2{\rm rk}\,G}$ outside the integral. This can be justified by noting that at $\alpha=0$ one obtains only this factor which can then be interpreted as the Schur index of $U(1)^{{\rm rk}\,G}$ vector multiplets. This is exactly as expected from the $p\,q=t$ first limit. As is immediately obvious, setting $\alpha=1$ reduces \eref{eq:NoNorm_GenSchur_def} to the usual gauge theory Schur index.

One can now ask how to introduce the 4d central charge dependant overall $q$-factor in this expression. This is the same as writing the analogue of the generalised Schur partition function corresponding to \eref{eq:Lag_SchurIndex}. We propose the following {\it normalised} \footnote{Note that our usage of ``normalised'' here refers to the $q^{c_{\rm 4d}/2}$ factor, and not the numerical factor that was already defined in \eref{eq:NoNorm_GenSchur_def} such that the leading order coefficient is 1.} generalised Schur partition function as the one with the proper inclusion of the 4d central charge
\begin{equation}\label{eq:GenSchur_modular}
    \begin{split}
        &\cZ_G(q;\alpha) \\
        &\quad = C(\alpha)\,\frac{\eta(\tau)^{2\,{\rm rk}\,G}}{|W|} \oint\prod_{j=1}^{{\rm rk}\,G} \frac{{\rm d}z_j}{2\pi iz_j} \bigg( \eta(\tau)^{{\rm rk}\,G-{\rm dim}\,G}\prod_{\bm\alpha\neq0} \vartheta_1(\bm\alpha\cdot\mbf u,\tau) \prod_l\prod_{\rho\in\mathcal R_l} \frac{\eta(\tau)^{N_l}}{\vartheta_4^{N_l}(\bm\rho\cdot\mbf u),\tau)} \bigg)^\alpha ~,
    \end{split}
\end{equation}
where $C(\alpha)\in\mathbb C$ is an overall normalisation defined such that the leading order coefficient is normalised to 1. In this case, it can simply be expressed as
\begin{equation}
    C(\alpha) = \frac{(-i)^{{\rm rk}\,G-{\rm dim}\,G}}{N(\alpha)} ~.
\end{equation}

This $\alpha$-dependent interpolating limit of the superconformal index is well-defined for general $\alpha\in\mathbb R^+$ and was studied in \cite{Deb:2025ypl}. Generically, the generalised Schur partition function yields a $q$-series with non-integer coefficients. However, for finely-tuned integral/rational values of $\alpha$, one obtains a $q$-series with integer coefficients and some cases were identified with the Schur index of different 4d $\cN=2$ SCFTs for certain special values of $\alpha$.

\subsection{Modular differential equations and the indicial equation}
\label{subsec:MLDEreview}

It has long been known \cite{Anderson:1987ge,Mathur:1988na,Eguchi:1988wh} that the characters of 2d Rational Conformal Field Theories (RCFT) satisfy Modular Linear Differential Equations (MLDE). Over the last few decades, this has been used with some success to classify RCFT (see for example \cite{Mukhi:2022bte} and references therein). MLDEs are characterised by their Wronskian index $\ell$, a non-negative integer labelling the number of zeroes of the leading Wronskian determinant. 

The general form of an MLDE is:
\be
\Big(\cD^n+\sum_{k=1}^n\sum_i \mu_{k,i}\,\phi_{2k,i}(q)\cD^{n-k}\Big)\chi(q)=0
\label{genMLDE}
\ee
where $q=e^{2\pi i\tau}$ and $\cD$ is the Ramanujan-Serre derivative:
\be
\cD\equiv q\frac{d}{dq}-\frac{k}{6}E_2(q)
\label{RSderiv}
\ee
when it acts on a modular function of weight $2k$. Here $\mu_k$ are constants and $\phi_{2k,i}(q)$ are meromorphic modular functions for $SL(2,Z)$ of weight $2k$, with $i$ labelling the independent functions for the given weight. The maximum number of poles they can have is given by the Wronskian index. We always normalise the $\phi_{2k,i}$ so that their $q$-series starts with 1. In the special case $\ell=0$ the $\phi_{2k,i}$ are holomorphic and hence modular {\em forms} that belong to the ring generated by the Eisenstein series $E_4(\tau),E_6(\tau)$. Even in this simple case, the number of parameters in the equation grows quadratically with $n$, making it increasingly hard to solve this equation. 

One of the useful pieces of information that can be extracted from an MLDE is the indicial equation which we now describe. For this we note that the general solution (except in cases where there are factors of $\log q$, which we temporarily ignore) takes the power-series form:
\be
\chi_A(q)=q^{\gamma_A}\sum_{r=0}^\infty a_{A,r}\,q^r
\label{charq}
\ee
where the $a_{A,r}$ are constant coefficients and the $\gamma_A$, known as critical exponents, are determined in terms of the coefficients $\mu_{k,i}$. For MLDEs whose solutions are the characters of an RCFT, we identify the exponents with the conformal data $c,h_A$ by:
\be
\gamma_0=-\frac{c}{24},\qquad \gamma_A=\gamma_0+h_A,~~A=1,2,\cdots,n-1
\ee

There is an important ambiguity inherent in the above association. MLDEs simply provide a set of solutions, each with its own critical exponent, but no a priori guidance on which one to select as $\gamma_0$. However if we require the result to be a {\em unitary} VOA then we are obliged to select the lowest exponent as $\gamma_0$ which fixes the above ambiguity. This choice ensures that all the $h_i$ are positive, a necessary (though not sufficient) condition for unitarity. We call this the ``unitary presentation'' of the characters.

Without unitarity, in principle the same set of characters can describe different VOAs, and several examples are known. For example with $n=2$ and $\ell=0$, for a suitable choice of the parameter, the MLDE has a pair of solutions with exponents $\left(-\frac{1}{6}, \frac{1}{3}\right)$. In the unitary presentation the first one is the identity character and we therefore have $c=4, h=\frac12$. This corresponds to the $SO(8)_1$ VOA. However if we reverse the presentation then the new identity character is the second one and hence $c=-8$. This is known to be a character of $V_{-\frac32}(SU(3))$.

A useful formula relates the central charge before and after exchanging presentations. If we start with a definite presentation labelled by $(c,h_1,h_2,\cdots,h_{n-1})$ and then choose the exponent $-\frac{c}{24}+h_i$ to be the identity instead, we easily find:
\be
(c,h_1,h_2,\cdots,h_i,\cdots,h_{n-1})\to (c-24h_i, h_1-h_i,h_2-h_i,\cdots, -h_i, \cdots, h_{n-1}-h_i)
\label{presexchange}
\ee

The Wronskian index for the MLDE is the number of zeroes of a Wronskian determinant made from the solutions $\chi_A$. For any fixed $\ell$, the exponents satisfy the following equation:
\be
\sum_{A=0}^{n-1}\gamma_A=-\frac{n\,c}{24}+\sum_{A=1}^{n-1}h_A=\frac{n(n-1)}{12}-\frac{\ell}{6}
\label{RiemRoch}
\ee

If we know the exponents $\gamma_A$ we can determine the quantities $\nu_k=\sum_i\mu_{k,i}$ in the MLDE, though not the individual values of $\mu_{k,i}$. To see this let us count the number of parameters on both sides. On one side the number of exponents  is $n$, while on the other hand the MLDE has a constant $\mu_{k,i}$ for every independent modular form of weight  $2k$. Since $k$ takes $n$ values, there are $n$ modular weights $2k=2,4,6,\cdots,2n$ in the sum. Now the exponents are constrained by \eref{RiemRoch}, but the coefficient $\nu_1$ turns out to be constrained by the same equation as we will see below. Thus given the exponents, we can determine the $n-1$ independent quantities $\nu_k,k=2,3,\cdots,n $. Except at low weights, there are multiple modular forms for a given weight, which is the reason for the additional label $i$. Clearly knowing the $\nu_k$ is not enough to determine the all the constants $\mu_{k,i}$ and hence the equation. Rather, for a given set of exponents there are multi-parameter families of distinct MLDE (with distinct solutions) obtained by varying the individual $\mu_{k,i}$ keeping their sums $\nu_k$ fixed.

We now work out the indicial equation that relates the $\nu_k$ to the $\gamma_A$. For this, we insert the leading power in \eref{charq} into \eref{genMLDE} and retain the leading power of the result. This then has to be set to 0 to satisfy the MLDE in leading order. We see that $q\frac{d}{dq}$ on $q^\gamma$ gives back $\gamma\,q^\gamma$, so we can replace the former operator by the number $\gamma$. However the derivatives $\cD$ in the MLDE are not simply $q\frac{d}{dq}$, but contain factors of Eisenstein $E_2$. Since each action of $\cD$ on a modular function of fixed weight increases its weight by 2, we have:
\be
\cD^r =\prod_{j=r-1}^{0}\left(q\frac{d}{dq}-\frac{j}{6}E_2(q)\right)
\ee
where the terms are ordered with $j$ decreasing towards the right. Applying this operator to $q^\gamma$, we find terms where every derivative acts on $q^\gamma$ and gives back $\gamma\,q^\gamma$, but also terms where some of the derivatives act on an $E_2$. However the latter have sub-leading (higher) powers of $q$, so they can be ignored as long as we are only keeping the leading power. Thus we may write:
\be
\cD^r \left(q^\gamma\right)\sim \prod_{j=0}^{r-1}\left(\gamma-\frac{j}{6}\right)q^\gamma
\ee
To leading order we can also replace each $\phi_{2k,i}$ by its leading term, which has been normalised to 1. Thus we get the indicial equation:
\be
\prod_{j=0}^{n-1}\left(\gamma-\frac{j}{6}\right)+\sum_{k=1}^{n}\nu_k\prod_{r=0}^{n-k-1}\left(\gamma-\frac{r}{6}\right)=0
\label{indicial}
\ee
where we used the definition $\nu_k=\sum_i\mu_{k,i}$. Expanding and retaining the highest powers, we obtain the form:
\be
\gamma^n-\frac{n(n-1)}{12}\gamma^{n-1}+\nu_1\gamma^{n-1}+\cdots=0
\label{indicialfirsttwo}
\ee
where $\cdots$ denotes terms with lower powers of $\gamma$. Now since the leading power of \eref{indicialfirsttwo} has coefficient unity, the LHS of \eref{indicial} can equivalently be written:
\be
P_n(\gamma_A):=\prod_{A=0}^{n-1}(\gamma-\gamma_A)=0
\ee
where $\gamma_A$ are the roots. Equating the power $\gamma^{n-1}$ between the two forms, we find:
\be
\sum_{A=0}^{n-1} \gamma_A=\frac{n(n-1)}{12}-\nu_1
\ee
which we recognise as \eref{RiemRoch} with $\nu_1=\frac{\ell}{6}$. Thus $\nu_1$ is determined by the Wronskian index. Thereafter, \eref{indicial} can be used to determine the exponents $\gamma_A$ in terms of the independent parameters $\nu_2,\nu_3,\cdots\nu_n$, and vice-versa. In fact, determining $\nu_k$ from $\gamma_A$ is the simpler direction, as we just have to solve linear simultaneous equations.

 As an example, for $n=3$ and $\ell=0$ we get:
\be
\gamma^3-\half \gamma^2+\left(\nu_2+\frac{1}{18}\right)\gamma+\nu_3=0
\label{indicial.neq3}
\ee
which tells us that \cite{Mathur:1988gt}:
\be
\begin{split}
\nu_2 &= e_2(\gamma_A)-\frac{1}{18}\\
\nu_3 &= -e_3(\gamma_A)
\end{split}
\ee
with $e_k$ being the elementary symmetric polynomials of degree $k$.

\subsection{Contour integral representation for RCFT characters}
\label{subsec:contourintegral_review}

Many of the analyses of MLDE and their solutions have focused on the simplest case, $\ell=0$. However even in this case, it becomes increasingly difficult to explicitly solve the equations beyond working order-by-order in a $q$-series by the Frobenius method. In particular, a closed-form solution is required in order to compute the modular transformations of the solutions and this is not generally available when the number of independent characters $n$ (closely related to the number of primaries of the full chiral algebra) is large. This contrasts with the case of two characters where the MLDEs  are solved \cite{Mathur:1988gt, Naculich:1988xv} by hypergeometric functions whose modular transformations are known.

To redress this problem, a multi-variable contour integral form was put forward in \cite{Mukhi:1989qk} that is capable of solving MLDEs after they are re-expressed in terms of the modular $\la$-function (defined below), even for arbitrarily large numbers of characters. In the above reference it was shown that the contour integrals describe the characters of several infinite families of RCFTs, including the $c<1$ minimal models as well as the $SU(2)_k$ and $SU(N)_1$ WZW models. This was somewhat surprising as the contour integrals, which we review below, depend on just two independent real constants -- so it is not guaranteed that every RCFT can be represented by such integrals. Aspects of the contour-integral proposal for characters were worked out in more detail in \cite{Mukhi:2019cpu} and an algorithm found to explicitly compute the modular $\mathcal S$-matrix for these contour integrals.

The contour integral approach of \cite{Mukhi:1989qk} was adapted from the Dotsenko-Fateev integrals used in \cite{Dotsenko:1984ad} to solve for the 4-point correlators of minimal model CFT. These in turn are adapted from Aomoto-Selberg integrals. It was noted in \cite{Mukhi:1989qk} that if the cross-ratio $z$ for correlators is replaced by the modular $\la$-function:
\be
\la=\frac{\vartheta_2^4(\tau)}{\vartheta_3^4(\tau)}
\ee
then the BPZ differential equation for correlators turns into the MLDE for characters, expressed in terms of $\la$ \footnote{This use of the $\la$ function was also studied more recently in \cite{Cheng:2020srs}.}. This in turn implies that the solutions of the BPZ equation in terms of Dotsenko-Fateev contour integrals can be mapped on to solutions of MLDEs, providing a formalism in which MLDEs can be solved explicitly without resorting to a $q$-expansion. The calculation of monodromies for conformal blocks then nicely turns into a calculation of the modular $S$-matrix for characters.

The contour integrals in question are multiple integrals over variables $t_i, i=1,2,\cdots,n-1$ \footnote{There is a generalisation to two families of variables \cite{Dotsenko:1984ad,Mukhi:1989qk} that we will not need in the present work.}. They are given by:
\be
J_{A}(a,b|\la) = \cN_{A}\big(\la(1-\la)\big)^p
\int_{\cC_A} \prod_{i=1}^{n-1}dt_i\,\Big[  t_i\,(t_i-1)(t_i-\la)\Big]^a \prod_{1\le j < i \le n-1}(t_i-t_j)^b
\label{JAp}
\ee
with $0\le A\le n-1$. Here $a,b$ are the only free parameters, while $p$ is determined in terms of them as:
\be
p = -\frac{n-1}{3}\Big(1+3a +\frac{b}{2}(n-2)\Big)
\label{prefac}
\ee
The constants $\cN_{A}$ are a set of normalisations that are worked out in detail in \cite{Mukhi:2019cpu}. The contour integrals are to be performed in the order $t_1, t_2,\cdots t_{n-1}$ and the subscript $\cC_A$ means the contours for $t_1,\cdots t_A$ run from 0 to $\la$, while those for $t_{A+1},\cdots t_{n-1}$ run from 1 to $\infty$. In particular, this means that for $A=n-1$ (the maximum allowed value), all contours run from 0 to $\la$.

Now $\la$ is a Hauptmodul of the subgroup $\Gamma(2)\subset SL(2,Z)$, and it transforms under $SL(2,Z)$ as:
\be
S{}:\quad\la \to 1-\la, \qquad T{}:\quad\la\to \frac{\la}{\la-1}
\ee
Using these one can show that the contour integrals \eref{JAp} transform into a linear combination of themselves under $SL(2,Z)$. This is a key feature that allows them to potentially describe the characters of an RCFT.

The simplest version of the above family of integrals is the case $n=2$. In this case there are no terms of the type $\prod(t_i-t_j)$ in the integral, so $b$ drops out and we get the one-parameter family:
\be
\begin{split}
J_{0} & = \cN_{0} \big(\la(1-\la)\big)^{-a-\frac13}
\int_1^\infty dt\,\Big[t(t-1)(t-\la)\Big]^a\\
J_{1} & = \cN_{1} \big(\la(1-\la)\big)^{-a-\frac13}
\int_0^\la dt\,\Big[t(t-1)(t-\la)\Big]^a
\end{split}
\label{twochar.0}
\ee
For small $\la$, we have $\la\simeq \sqrt{q}$. Hence the leading behaviour of the two integrals above is:
\be
\begin{split}
J_{0} &~\sim~ \la^{-a-\frac13}~\sim~ q^{-\frac{a}{2}-\frac16}\\
J_1 &~\sim~ \la^{a+\frac23} ~\sim~ q^{\frac{a}{2}+\frac13}
\end{split}
\ee
These can be identified with a pair of characters of a two-character RCFT of central charge $c$ whose non-trivial primary has holomorphic conformal dimension $h$, as follows. We equate the leading behaviours using:
\be
\begin{split}
-\frac{a}{2}-\frac16 &=-\frac{c}{24}\\
\frac{a}{2}+\frac13 &=-\frac{c}{24}+h
\end{split}
\label{expmatch}
\ee
This implies:
\be
-\frac{c}{12}+h=\frac16
\label{RRtwo}
\ee
Comparing with the general Riemann-Roch formula for $n$ characters \eref{RiemRoch}, we find $\ell=0$. Now, any pair of functions transforming into themselves under $SL(2,Z)$ and having $\ell=0$ satisfies the MMS equation \cite{Mathur:1988na}:
\be
\Big(\cD^2+\mu\,E_4(\tau)\Big)\chi=0
\label{MMSeq}
\ee
where $\mu=-\frac{c(c+4)}{576}$. Thus the pair \eref{twochar.0} satisfies \eref{MMSeq} for any value of the parameter $a$, with the following identifications arising from \eref{expmatch}: 
\be
\begin{split}
c &=12a+4\\
h &=a+\half\\
\mu &=-\frac{(a+\frac13)(a+\frac23)}{4}
\end{split}
\label{MMSchmu}
\ee
For specific values of $a$ the contour integrals are the characters of the entire MMS series CFTs \cite{Mathur:1988na, Mathur:1988gt} as well as infinitely many additional vector-valued modular forms that have integral but not necessarily positive $q$-coefficients \cite{Chandra:2018pjq}, known as quasi-characters, and also solutions having logarithmic behaviour in $q$. 

Returning now to the contour integral \eref{JAp} for general $n$, it is easily shown that if it corresponds to the characters of an RCFT, the latter will have $\ell=0$. This is done by noting that the leading behaviour of each integral at small $\la$ is:
\be
J_A(\la)~~{\buildrel\la\to 0 \over \sim}~~\la^{p+\Delta_A}  ~\sim~q^{\half(p+\Delta_A)} 
\ee
where:
\be
\Delta_A = A(1+2a)+\half b\,A(A-1),\quad A=1,2,\cdots,n-1
\label{Deltadef}
\ee
Identifying the leading exponents $\half(p+\Delta_A)$ with $-\frac{c}{24}+h_A$, we find that the central charge and conformal dimensions are:
\be
\begin{split}
c &= -12p = 4(n-1)\left(1+3a+\frac{b}{2}(n-2)\right)\\
h_A &= \frac{\Delta_A}{2} = A\left(a+\half\right)+\frac{b}{4}A(A-1)
\end{split}
\label{chident}
\ee
Applying \eref{RiemRoch} then gives $\ell=0$ \cite{Mukhi:1989qk}. 

This tempts us to ask under what conditions the contour integral construction above describes the characters of RCFTs. A key advantage of doing so is that the modular $S$-matrix can be worked out explicitly, as explained in detail in \cite{Mukhi:2019cpu} generalising the analogous construction for the monodromy of conformal blocks in \cite{Dotsenko:1984ad}. To see when this can be done, let us assume a given $\ell=0$ RCFT with data $c,h_A$ and ask whether its characters can be written as contour integrals as above. For this one needs to identify the leading exponents $-\frac{c}{24}+h_A$ of the RCFT with the leading exponents $\half(p+\Delta_A)$ of the contour integrals. But the latter set depends on two parameters $a,b$ for $n\ge 3$ and one parameter for $n=2$, while the former set has $n-1$ independent parameters after imposing $\ell=0$. Thus, identification is assured only for $n=2,3$ where the number of parameters is equal, while there will be RCFTs with $n\ge 4$ characters that cannot be identified with the above contour integrals. However, \cite{Mukhi:1989qk} noted the surprising fact that several well-known infinite families of RCFTs admit an identification of their exponents with the leading $q$-behaviour of the contour integrals. This was shown to hold for the characters of the $A$, $D$ and $E$ series $SU(2)_k$ models as well as for the $SU(N)_1$ WZW models and the $(2,2p+1)$ BPZ minimal models. Additionally, by generalising \eref{JAp} to two sets of integration variables (Eq. 2.12 of \cite{Mukhi:1989qk}), matching of exponents was achieved for the A, D and E series characters of all $(p,p')$ BPZ minimal models.

However matching of exponents is merely a necessary, but not sufficient, condition to identify the characters themselves with the contour integrals. But for $\ell=0$, the exponents completely determine the MLDE for $n\le 5$, therefore in this case the necessary condition is also sufficient. On the other hand for $n>5$, even after successfully matching exponents, we have no guarantee that the contour integral matches a given RCFT. In Ref. \cite{Mukhi:1989qk} it was conjectured that in a large number of cases where the RCFT exponents match those of contour integrals, the latter indeed represent the characters of the RCFT. Non-trivial evidence was provided for this conjecture. However it has not been proved in generality as far as we know, nor is the complete list of RCFT known for which this is true.

\section{$SU(2)$ SCFT with $4$ fundamental hypers}
\label{sec:su2_genSchur}

In this section, we consider the generalised Schur partition function associated to the 4d $\cN=2$ $SU(2)$ SCFT with 4 fundamental hypermultiplets. Since the fundamental representation of $SU(2)$ is pseudo-real, these can be considered as $8$ half-hypermultiplets, {\it i.e.}, chiral multiplets, in the fundamental representation of $SU(2)$. This is a rank one SCFT and also corresponds to theory of class $\mathcal S$ of type $SU(2)$ associated to a Riemann sphere with four maximal punctures.

The 2d vertex operator algebra associated to this SCFT is the simple affine vertex algebra $V_{-2}(SO(8))$ which has the central charge\footnote{Note that we use the notation $V_k(G)$ to refer to the simple affine vertex algbra associated to $G$ at level $k$ only when $k$ is negative or fractional. Otherwise we usually refer to it as $G_k$ which typically the denotes the corresponding Wess--Zumino--Witten model.}
\begin{equation}
    c_{\rm 4d} = -\frac{c_{\rm 2d}}{12} = -\frac{(-14)}{12} = \frac76 ~.
\end{equation}
By specialising \eref{eq:Lag_SchurIndex}, the Schur index can be expressed as the following integral taken over the unit circle $|z|=1$: 
\begin{equation}
    \begin{split}
        \cI_{SU(2)}(q) &= -\eta(\tau)^8 \oint \frac{{\rm d}z}{4\pi iz}\,\frac{\vartheta_1(2u,\tau)\vartheta_1(-2u,\tau)}{\vartheta_4^4(u,\tau)\,\vartheta_4^4(-u,\tau)} \\
        &= \eta(\tau)^8 \oint \frac{{\rm d}z}{4\pi iz}\,\frac{\vartheta_1^2(2u,\tau)}{\vartheta_4^8(u,\tau)} ~.
    \end{split}
\end{equation}
This Schur index, in fact, admits a closed form solution as a weight $6$ quasimodular form with a prefactor that only depends on the Dedekind eta functions \cite{Beem:2021zvt,Pan:2021mrw}
\begin{equation}
    \cI(q) = \frac{E_2(\tau)E_4(\tau)-E_6(\tau)}{720\,\eta(\tau)^{10}} ~.
\end{equation}
This enables an explicit computation of the two-dimensional vector-valued modular form associated to this Schur index as a closed form expression. This dimension of the VVMF is expected as the identity character of the corresponding VOA solves the following second order $\ell=0$ MLDE \cite{Mathur:1988na,KZ,KK},
\begin{equation}
    \left( \mathcal D^2 - \frac{35}{144} \,E_4(\tau) \right) \mathcal I_{SU(2)}(q) = 0 ~.
\end{equation}

The generalised Schur partition function associated to this theory was introduced in \cite{Deb:2025ypl} without the $q$-prefactor as a specialisation of \eref{eq:NoNorm_GenSchur_def}. Here, we want to study the normalised generalised Schur partition function as defined in \eref{eq:GenSchur_modular}
\begin{equation}\label{eq:SU2_genSchur}
    \cZ_{SU(2)}(q;\alpha) = \frac{\eta(\tau)^2}{N(\alpha)} \oint \frac{{\rm d}z}{4\pi iz}\, \left( \eta(\tau)^6\frac{\vartheta_1^2(2u,\tau)}{\vartheta_4^8(u,\tau)} \right)^\alpha ~.
\end{equation}
It was observed in \cite{Deb:2025ypl} that for special rational values of $\alpha$, $\cZ_{SU(2)}(q;\alpha)$ reproduces the Schur indices of the of rank one SCFTs that correspond to the non-unitary Deligne--Cvitanovi\'c series of VOAs. All of these Schur indices, or equivalently the vacuum characters of the associated VOAs, are well known to satisfy the second order $\ell=0$ MLDE.

This suggests a deeper relationship between the generalised Schur partition function and the vector-valued modular forms that, for special values of a certain parameter, correspond to the characters of two-character RCFTs with vanishing Wronskian index $\ell=0$. We now show that this is indeed the case, by proving that \eref{eq:SU2_genSchur} is equivalent to the contour integral representation of the general solution of the second order $\ell=0$ MLDE which, in particular, is satisfied by all 2d RCFT  with two characters and $\ell=0$.

\begin{prop}\label{prop:SU2}
    The generalised Schur partition function $\cZ_{SU(2)}(q;\alpha)$, with analytically continued $\alpha\in\mathbb C$, is a solution to the following second order $\ell=0$ modular linear differential equation, the MMS equation:
    \begin{equation}\label{eq:MMS_equation}
        \left(\mathcal D^2 + \mu \,E_4(\tau)\right)\,\cZ_{SU(2)}(q;\alpha) = 0 \;,\; \text{ with } \mu = -\frac{(6\alpha-1)(6\alpha+1)}{144} ~.
    \end{equation}
\end{prop}

\noindent{\bf Proof} : We prove this proposition by mapping the generalised Schur partition function $\cZ_{SU(2)}(q;\alpha)$ to the contour integral representation \eref{twochar.0}. To make the notation less cumbmersome, we define the following shorthand notation for the Jacobi theta functions
\begin{equation}\label{eq:Compact_Not}
    \vartheta_i(u) \equiv \vartheta_i(u,\tau) \;,\; \vartheta_i \equiv \vartheta_i(0) \equiv \vartheta_i(0,\tau) \;,\; \eta \equiv \eta(\tau) ~,
\end{equation}
where we also simplify the notation for the Dedekind $\eta(\tau)$ for consistency. Moreover, since we want to show that $\cZ_{SU(2)}(q;\alpha)$ satisfies an MLDE, any $q$-independent constant factor can be ignored due to the linearity of the differential equation. Therefore, we will drop such constant terms throughout this proof as they are irrelevant for the final result.

Let us now focus on the original integrand of $\cZ_{SU(2)}(q;\alpha)$, {\it i.e.},
\begin{equation}\label{eq:SU2_integrand}
    I_{SU(2)}(q,z) = \eta^6\,\frac{\vartheta_1^2(2u)}{\vartheta_4^8(u)} ~,
\end{equation}
where we recall that $z=e^{2\pi iu}$. One can now use the theta doubling identity \eref{theta1.addition.identity} for $\vartheta_1(2u)$,
\begin{equation}
    \vartheta_1(2u,\tau) = 2\,\frac{\vartheta_1(u)\vartheta_2(u)\vartheta_3(u)\vartheta_4(u)}{\vartheta_2\vartheta_3\vartheta_4} ~,
\end{equation}
to simplify the integrand to
\begin{equation}
    I_{SU(2)}(q,z) = \frac{4\,\eta^6}{\vartheta_2^2\,\vartheta_3^2\,\vartheta_4^2}\left(\frac{\vartheta_1(u)\,\vartheta_2(u)\,\vartheta_3(u)}{\vartheta_4^3(u)}\right)^2 = \left(\frac{\vartheta_1(u)\,\vartheta_2(u)\,\vartheta_3(u)}{\vartheta_4^3(u)}\right)^2 ~,
\end{equation}
where in the second step we use \eref{theta.eta} to simplify the $z$-independent factor. Using the Jacobi elliptic functions defined in Appendix \ref{subsec:Jacobi_elliptic}, we can rewrite this as\footnote{There are several notational conventions for Jacobi elliptic functions: ${\rm sn}(v,k^2), {\rm sn}(v,\lambda), {\rm sn}(v,k)$. We choose the last one to make the notation less cumbersome while keeping in mind that $k^2=\lambda$.}
\begin{equation}\label{eq:SU2_integrand.2}
    I_{SU(2)}(q,z) = \frac{\vartheta_2^4}{\vartheta_4^4}\,{\rm sn}^2(v,k)\,{\rm cn}^2(v,k)\,{\rm dn}^2(v,k) ~,
\end{equation}
where $k\equiv k(\tau)$ is the ellpitic modulus defined in \eref{eq:elliptic_modulus} and 
\begin{equation}
    v \coloneq \pi\,\vartheta_3^2\,u ~,
\end{equation}
is a rescaling of the $u$ parameter.

We can now define a new variable $t$ as
\begin{equation}\label{eq:var_change}
    t \coloneq \lambda\,{\rm sn}^2(v,k) ~,
\end{equation}
which implies that
\begin{equation}
    {\rm cn}^2(v,k) = 1 - {\rm sn}^2(v,k) = 1 - \frac{t}{\lambda} \;,\; {\rm dn}^2(v,k) = 1 - \lambda\,{\rm sn}^2(v,k) = 1 - t ~.
\end{equation}
Together with the simple relation
\begin{equation}
    \frac{\vartheta_2^4}{\vartheta_4^4} = \frac{\lambda}{1-\lambda} ~,
\end{equation}
we can simplify the integrand to
\begin{equation}
    I_{SU(2)}(q,z) = \frac{t(1-t)(\lambda-t)}{\lambda(1-\lambda)} ~.
\end{equation}
Thus we see that the original integrand \eref{eq:SU2_integrand} of the generalised Schur partition function $\cZ_{SU(2)}(q;\alpha)$, can be naturally reduced a simple polynomial in the variable $t$ related to the Jacobi elliptic functions.

Let us now get back to the full integral to see the full effect of the variable change in \eref{eq:var_change}. Recall, that we have performed a chain of variable changes
\begin{equation}
    z \;\longrightarrow\; u = \frac{1}{2\pi i}\,{\rm log}z \;\longrightarrow\; v = \pi\,\vartheta_3^2\,u \;\longrightarrow\; t = \lambda\,{\rm sn}^2(v,k) ~.
\end{equation}
Therefore, the initial $SU(2)$ measure transforms as
\begin{equation}\label{eq:Jacobian}
    \frac{{\rm d}z}{2\pi i z} = {\rm d}u = \frac{1}{\pi\,\vartheta_3^2}\,{\rm d}v = \frac{1}{2\pi\,\vartheta_3^2}\,\frac{{\rm d}t}{\sqrt{t(1-t)(\lambda-t)}} ~,
\end{equation}
where for the final transformation we use \eref{sn.derivative}. Therefore, the generalised Schur partition function at this stage can be expressed as
\begin{equation}\label{eq:SU2_temp_nocontour}
   \begin{split}
        \cZ_{SU(2)}(q;\alpha) &\propto \frac{\eta(\tau)^2}{\vartheta_3(\tau)^2}\,\left[\lambda(1-\lambda)\right]^{-\alpha} \int_{\mathcal C}{\rm d}t\,\Big[t(1-t)(\lambda-t)\Big]^{\alpha-\frac12} ~, \\
        &\propto\left[\lambda(1-\lambda)\right]^{-\alpha+\frac16} \int_{\mathcal C_1}{\rm d}t\,\Big[t(1-t)(\lambda-t)\Big]^{\alpha-\frac12} ~.
    \end{split}
\end{equation}
where we have not yet specified $\mathcal C_1$, and we used \eref{theta.eta} once again to simplify the $\eta^2/\vartheta_3^2$ factor.

Let us now discuss the final piece of this proof: the contour $\mathcal C_1$. The original integral over the unit circle $|z|=1$ corresponds to integrating over the $u$ variable in the range $[0,1)$. Therefore, since the variable $v$ is a simple rescaling of $u$, we have that
\begin{equation}
    v \in [0,2K(k)) \;,\; \text{ where } 2K(k) = \pi\,\vartheta_3^2 ~,
\end{equation}
is also known as the complete elliptic integral of the first kind, as reviewed in Appendix \ref{subsec:Jacobi_elliptic}. The last transformation is highly non-linear and {\it a priori} the form of $\mathcal C$ is not clear. However, a massive simplification occurs due to the periodicity of ${\rm sn}^2(v,k)$. In particular, it can be shown that
\begin{equation}
    {\rm sn}(0,k) = -{\rm sn}(2K(k),k) = 0 \;,\; {\rm sn}(K(k),k) = 1 ~.
\end{equation}
This implies that as the $v$ contour runs from $0$ to $2K(k)$, the contour in the $t$-plane runs from $0$ to $\lambda$ and then back to $0$.

Due to the analytic continuation of $\alpha$ to $\mathbb C$, the integral in \eref{eq:SU2_temp_nocontour} has branch points at $t=0,\lambda,1,\infty$. Thus we can choose two branch cuts that run from $0$ to $\lambda$ and from $1$ to $\infty$. This implies that the contour in the $t$-plane loops around one of these branch cuts in the clockwise direction. However by a standard manipulation this closed contour can be split into 4 pieces: (1) a piece in the upper half plane from $0$ to $\lambda$, (2) a semi-circle around $\lambda$, (3) a piece in the lower half plane from $\lambda$ back to 0, and (4) a semi-circle around $0$. Due to the branch cut, the semicircles pick up an $\alpha$-dependent phase, while piece (3) can be identified with (1) up to a sign. The end result is that this closed contour is proportional to the contour that runs from $0$ to $\lambda$.

Therefore, the generalised Schur partition function can finally be expressed as
\begin{equation}
    \cZ_{SU(2)}(q;\alpha) \propto \left[\lambda(1-\lambda)\right]^{-\alpha+\frac16} \int_0^\lambda{\rm d}t\,\Big[t(1-t)(\lambda-t)\Big]^{\alpha-\frac12} ~.
\end{equation}
It is easy to observe now that this is the same as $J_1$ in \eref{twochar.0} for
\begin{equation}
    a = \alpha - \frac12 ~.
\end{equation}
Therefore, as shown in \cite{Mukhi:1989qk} and reviewed in Section \ref{subsec:contourintegral_review}, $\cZ_{SU(2)}(q;\alpha)$ satisfies the MMS equation (second order $\ell=0$ MLDE) of \eref{MMSeq} with:
\begin{equation}
    \mu = -\frac{(6\alpha-1)(6\alpha+1)}{144},\quad c=12\alpha-2,\quad h=\alpha ~.
\end{equation}
This concludes the proof of Proposition \ref{prop:SU2}.
$\hfill\square$

From the contour we see that $\cZ_{SU(2)}(q;\alpha)$ corresponds to the non-identity character $J_1$ rather than the identity character. Therefore, in terms of the conformal dimension $h$ of the primary of a putative two-character unitary RCFT, we can say that $\cZ_{SU(2)}(q;\alpha)$ for a given $\alpha$ agrees with the non-identity character of the CFT, if we choose $\alpha=h$. We Refer to Appendix \ref{app:general_solution_MMS} for details about the general solutions to the MMS equation.

It is well-known that sending $h\to -h$ exchanges the role of the two characters, hence we can equally well say that the generalised Schur partition function with $\alpha=-h$ agrees with the identity character of some (non-unitary) RCFT. With this second interpretation, positive $\alpha$ is equated to negative $h$, and for $\alpha=1$ we therefore recover the familiar statement that the Schur index of a unitary 4d theory matches the identity character of a non-unitary quasi-lisse vertex operator algebra, corresponding in many cases of rank-one SCFTs to affine vertex algebras at negative fractional levels \cite{Arakawa:2016hkg, Beem:2017ooy}. For the actual Schur index this is the natural matching, because the Schur limit of the 4d theory reproduces not just the characters but an actual non-unitary 2d vertex operator algebra \cite{Beem:2013sza} (including examples involving logarithmic CFTs whose identity character is a quasimodular form \cite{Beem:2021zvt,Pan:2021mrw}). However for the generalised Schur partition function one can consider both interpretations of this partition function: as the second (non-identity) character of a unitary RCFT, or as the first (identity) character of a non-unitary one. Both interpretations agree with our Proposition \ref{prop:SU2}.

Finally, Proposition \ref{prop:SU2} not only proves the numerical observations made for the generalised Schur partition function of the $SU(2)$ SCFT, but also provides an explanation in the form of the MLDE as the hidden underlying structure. This also yields an alternate explanation for why $\cZ_{SU(2)}(q;\alpha)$ only produces some of the rank-one SCFTs, while the Schur index of some other rank-one SCFTs, such as the $SU(2)$ $\cN=4$ SYM, does not appear for any value of $\alpha$ -- the reason being that it does not satisfy the MMS equation \cite{Beem:2021zvt}. We also stress that Proposition \ref{prop:SU2} implies that $\cZ_{SU(2)}(q;\alpha)$ generates the whole zoo of solutions to the MMS equation: vacuum characters of non-unitary VOAs associated to rank-one SCFTs, both characters of unitary 2d rational conformal field theories (depending on the choice $\alpha=\pm h$), infinite families of quasi-characters, and logarithmic characters. 

The above observations in particular unify the results in Table 1 of \cite{Deb:2025ypl}, which lists the values of $\alpha$ that lead to the Schur index of some 4d theory. The corresponding non-unitary Deligne-Cvitanovi\'c CFT in every case has a central charge $c=-12c_{4d}$ and a conformal dimension $h=-\alpha$. Either one of these data, when converted to a value of $\mu$ from \eref{MMSchmu} and inserted into \eref{MMSeq}, gives rise to a CFT whose identity character corresponds to the Schur index. At the same time, if we choose  instead the identification $\alpha=h$, the same procedure leads to unitary MMS theories for which the {\em non-identity} character matches the Schur index (for the last 5 entries of that Table), or logarithmic theories with the same property (for the first 6 entries of the Table). 

Let us describe an example to see how the two alternate descriptions work. The $SU(3)_1$ RCFT is a member of the series of unitary MMS theories. It has $(c,h)=(2,\frac13)$ and its identity character has the $q$-expansion:
\be
\chi_0(q)= q^{-\frac{1}{12}}\left(1+8q+\cO(q^2)\right)
\ee
The coefficient 8 of the second term is just the dimension of $SU(3)$, since first-level descendants of the identity state can only be created by $J_{-1}^a$ and cannot be null, therefore they count the dimension of the algebra. Now, the non-identity character of the same RCFT has the form:
\be
\chi_1(q)=q^{\frac{1}{4}}\left(3+9q+\cO(q^2)\right)
\label{A2nonid}
\ee
The 3 of the first term represents the ground state of the module which is the fundamental/anti-fundamental ($3/{\bar 3}$) of $SU(3)$. However if we divide throughout by 3, we find the striking result that the {\em same} $q$-series, after this change of normalisation, remains integral -- the coefficients of all $q$ powers are still integers. Then we can try to think of this series:
\be
{\tilde \chi}_0(q)=q^\frac14\left(1+3q+\cO(q^2)\right)
\ee
as the identity character of a different, non-unitary VOA where the current algebra has dimension 3. This of course should be $V_k(SU(2))$ for some level $k$. The next step is to seek a value of $k$ such that the theory has central charge $c-24h=-6$, which is the effective $c$ if we treat the second character as the vacuum. Using the standard formula for $SU(2)_k$:
\be
c=\frac{3k}{k+2}
\ee
and equating it to $-6$, we land on $k=-\frac43$. Thus one can reinterpret the character in \eref{A2nonid} as either the second/non-identity character of $SU(3)_1$, or the first/identity character of $V_{-\frac43}(SU(2))$. In the latter form, it has been shown to correspond to the Schur index of the $(A_1,A_3)$ Argyres-Douglas theory \cite{Buican:2015ina,Cordova:2015nma}. It should be kept in mind that this type manipulation of exchanging unitary and non-unitary characters does not necessarily lead to a meaningful interpretation in terms of RCFT/VOAs in every case.

\section{$USp(2N)$ SCFT with $2N+2$ fundamental hypers}
\label{sec:USpN_genSchur}

In this section, we consider the generalised Schur partition function associated to the the 4d $\cN=2$ $USp(2N)$ SCFT with $N_f=2N+2$ fundamental hypermultiplets. We carry out a series of manipulations, similar to those in the previous section, that will lead us to more general contour integrals of the form \eref{JAp} that describe the characters of 2d RCFTs with vanishing Wronskian index.

As in the case for $SU(2)\cong USp(2)$, the fundamental representation of $USp(2N)$ is psuedo-real and hence these can be counted as $4N+4$ half-hypermultiplets that are rotated by the $SO(4N+4)$ flavour symmetry. The rank of this SCFT is $N$, which is the total number of Casimir invariants of $USp(2N)$. This gauge theory also has a twisted class $\mathcal S$ realisation \cite{Gaiotto:2009hg,Gaiotto:2009we,Tachikawa:2009rb,Tachikawa:2010vg} and corresponds to the theory of type $SO(2N)$ associated to a Riemann surface with two untwisted maximal punctures $[1^{2N}]$ and two twisted minimal punctures \cite{Chacaltana:2013oka}. The central charge of this theory is simply given by
\begin{equation}\label{eq:central_usp2n}
    c_{\rm 4d} = \frac{2\,n_V+n_H}{12} = \frac{N(4N+3)}{6} ~,
\end{equation}
where recall that ${\rm dim}\,USp(2N) = N(2N+1)$.

Instead of writing the full Schur index directly, we split the various contributions in \eref{eq:Lag_SchurIndex} and simplify them independently for the reader's convenience. The contribution of the vectors for $USp(2N)$ can be explicitly written as
\begin{equation}
    \prod_{\bm\alpha\neq0} \vartheta_1(\bm\alpha\cdot\mbf u,\tau) = \prod_{i=1}^N \vartheta_1(\pm 2 u_i) \prod_{1\leq i<j\leq N} \vartheta_1(\pm u_i\pm u_j) ~.
\end{equation}
The remainder of this section will use the compact notation for the Jacobi theta and Dedekind eta functions introduced in \eref{eq:Compact_Not}. Meanwhile, the contribution from $2N+2$ fundamental hypermultiplets can be simplified in the following way:
\begin{equation}
    \begin{split}
        \prod_{k=1}^{2N+2} \prod_{i=1}^N \frac{\eta^2}{\vartheta_4(\pm u_i)} &= \eta(\tau)^{4N(N+1)} \prod_{i=1}^N \frac{1}{\vartheta_4^{2N+2}(\pm u_i)} \\
        &= \eta^{4N(N+1)} \prod_{i=1}^N \frac{1}{\vartheta_4^4(\pm u_i)} \prod_{1\leq i<j\leq N} \frac{1}{\vartheta_4^2(\pm u_i)\,\vartheta_4^2(\pm u_j)} ~.
    \end{split}
\end{equation}
Therefore, the Schur index for the $USp(2N)$ SCFT with $N_f=2N+2$ can be expressed as
\begin{equation}
    \begin{split}
        \cI_{USp(2N)}(q) = \frac{\eta^{2N(N+3)}}{2^N\,N!} \oint \prod_{j=1}^N \frac{{\rm d}z_j}{2\pi i z_j} \prod_{i=1}^N\frac{\vartheta_1^2(2 u_i)}{\vartheta_4^8(u_i)} \prod_{1\leq i<j\leq N} \frac{\vartheta_1^2(u_i\pm u_j)}{\vartheta_4^4(u_i)\,\vartheta_4^4(u_j)} ~,
    \end{split}
\end{equation}
where we use the identity $\vartheta_1(\pm u)=-\vartheta_1(u)^2$ and $\vartheta_4(\pm u)=\vartheta_4(u)^2$, and simplify the overall constant by noting that $(-1)^N(-i)^{-2N^2}=1$.

The (normalised) generalised Schur partition function of this theory can be expressed by using \eref{eq:GenSchur_modular}
\begin{equation}\label{eq:USpN_genSchur_1par}
    \begin{split}
        &\cZ_{USp(2N)}(q;\alpha) = \\
        &\quad \frac{\eta^{2N}}{|W|\,N(\alpha)} \oint \prod_{j=1}^N \frac{{\rm d}z_j}{2\pi i z_j} \bigg( \eta^{2N(N+2)} \prod_{i=1}^N\frac{\vartheta_1^2(2 u_i)}{\vartheta_4^8(u_i)} \prod_{1\leq i<j\leq N} \frac{\vartheta_1^2(u_i\pm u_j)}{\vartheta_4^4(u_i)\,\vartheta_4^4(u_j)} \bigg)^\alpha ~,
    \end{split}
\end{equation}
It was numerically observed in \cite{Deb:2025ypl} that for special rational values of $\alpha$, $\cZ_{USp(2N)}(q;\alpha)$ reproduces the Schur indices of of three families of type I Argyres--Douglas theories \cite{Argyres:1995jj,Argyres:1995xn,Xie:2012hs,Cecotti:2012jx,Cecotti:2013lda,Cordova:2015nma}: $D_2(\mf{sl}_{2N+1})$, $D_{2N+1}(\mf{sl}_2) \equiv (A_1,D_{2N+1})$, and $D_{2N+3}(\mf{sl}_2,[2]) \equiv (A_1,A_{2N})$.\footnote{Here we have followed the notation for Argyres--Douglas theories used in \cite{Beem:2023ofp}.} It can be numerically checked that the Schur indices of many of these theories satisfy an analogous order-($N+1$) $\ell=0$ MLDE. Furthermore, note that the $N=1$ case of the $USp(2N)$ SCFT is the $SU(2)$ with $N_f=4$ SCFT that we discussed in the previous section.

Building on the proposition in the last section and these empirical observations, we make the following proposition:
\begin{prop}\label{prop:USp2N}
    The generalised Schur partition function $\cZ_{USp(2N)}(q;\alpha)$, with analytically continued $\alpha\in\mathbb C$, is a solution to the following order-$(N+1)$ modular linear differential equation with vanishing Wronskian index:
    \begin{equation}\label{eq:USp2N_MLDE_1par}
        \left(\mathcal D^{N+1} + \sum_{k=2}^{N+1} \sum_i \mu_{k,i} \, \phi_{2k,i}(\tau)\, \mathcal D^{N+1-k}\right)\,\cZ_{USp(2N)}(q;\alpha) = 0 ~,
    \end{equation}
where $\sum_i \mu_{k,i}$ is determined by solving the indicial equation \eref{indicial} for this MLDE in terms of exponents $\gamma_A$ given by:
    \begin{equation}\label{MLDEcoeffs.2}
        \begin{split}
            \gamma_0 &= -\frac{N}{12}\Big(2(N+2)\alpha-1\Big) ~, \\[2mm]
            \gamma_A-\gamma_0 &= \frac{A(A+1)}{2}\alpha,\quad A=1,\cdots, N ~.
        \end{split}
    \end{equation}    
\end{prop}
While only the linear combinations $\nu_k=\sum_i\mu_{k,i}$ are determined by the exponents $\gamma_A$, in fact all the parameters of the MLDE are determined in terms of $\alpha$. To actually find them requires expanding the generalised Schur index in a $q$-series to a finite order and inserting it into the MLDE. In simple cases where there is a unique modular form for a given weight (true for MLDE of order $\le 5$) this extra step is of course unnecessary.

In the next two sections, we provide the proof for Proposition \ref{prop:USp2N}. 

\subsection{First non-trivial case: $N=2$}
\label{subsec:USp4_genSchur}

We will first focus on the generalised Schur partition function of $USp(4)$ with $6$ fundamental flavours\footnote{This is known to be dual to the $SU(2)$ gauging of the strongly-coupled $E_7$ Minahan--Nemeschansky theory \cite{Argyres:2007cn,Minahan:1996cj}} as the analysis for general $N$ mirrors that of $N=2$, and the $N=1$ case was already considered in the previous section. The generalised Schur partition function for this case takes the form
\begin{equation}
    \cZ_{USp(4)}(q;\alpha) = \frac{\eta^4}{8\,N(\alpha)} \oint \prod_{j=1}^2 \frac{{\rm d}z_j}{2\pi i z_j} \bigg( \eta^{16} \, \frac{\vartheta_1^2(2 u_1)\,\vartheta_1^2(2 u_2)}{\vartheta_4^8(u_1)\,\vartheta_4^8(u_2)} \frac{\vartheta_1^2(u_1\pm u_2)}{\vartheta_4^4(u_1)\,\vartheta_4^4(u_2)} \bigg)^\alpha ~,
\end{equation}
We now provide the proof of Proposition \ref{prop:USp2N} for $N=2$, and then discuss many special cases and the connection to unitary RCFTs.

\noindent{\bf Proof} :

The relevant integrand for this case can thus be expressed as
\begin{equation}
    I_{USp(4)}(q,z_{1.2}) = \eta^{16} \, \frac{\vartheta_1^2(2 u_1)\,\vartheta_1^2(2 u_2)}{\vartheta_4^8(u_1)\,\vartheta_4^8(u_2)} \frac{\vartheta_1^2(u_1\pm u_2)}{\vartheta_4^4(u_1)\,\vartheta_4^4(u_2)} ~.
\end{equation}
Let us do this piecewise, where the first piece in $I_{USp(4)}(q,z_{1,2})$ can be chosen to be
\begin{equation}
    I_{USp(4)}^{(1)}(q,z_{1,2}) = \eta^6\,\frac{\vartheta_1^2(2 u_1)}{\vartheta_4^8(u_1)} \times \eta^6\,\frac{\vartheta_1^2(2 u_2)}{\vartheta_4^8(u_2)} = I_{SU(2)}(q,z_1) \times I_{SU(2)}(q,z_2) ~,
\end{equation}
which can be immediately recognised as two copies of the integrand of the generalised Schur partition function of $SU(2)$. Therefore, we already have the result that under the redefinitions
\begin{equation}\label{eq:var_change_USp4}
    t_i = \lambda\,{\rm sn}^2(v_i,k) \;,\; \text{ where } v_i = \pi\,\vartheta_3^2\,u_i \;,\; \text{ for } i=1,2 ~,
\end{equation}
this simplifies to
\begin{equation}
    \begin{split}
        I_{USp(4)}^{(1)}(q,z_{1,2}) &= \frac{\vartheta_2^8}{\vartheta_4^8} \prod_{i=1}^2 \big[{\rm sn}^2(v_i,k)\,{\rm cn}^2(v_i,k)\,{\rm dn}^2(v_i,k)\big] \\
        &= \prod_{i=1}^2 \left[ \frac{t_i(1-t_i)(\lambda-t_i)}{\lambda(1-\lambda)} \right] ~.
    \end{split}
\end{equation}

This shows that the variable change \eref{eq:var_change_USp4}, which is the straightforward multi-variable generalisation of \eref{eq:var_change}, is quite natural for the first piece of the $USp(4)$ integrand, $I_{USp(4)}^{(1)}(q,z_{1,2})$. A natural question is how the second piece,
\begin{equation}
    I_{USp(4)}^{(2)}(q,z_{1,2}) = \eta^4 \left( \frac{\vartheta_1(u_1\pm u_2)}{\vartheta_4(u_1)^2\,\vartheta_4(u_2)^2} \right)^2 ~,
\end{equation}
behaves under this elliptic variable change. To address this, we first note the ``theta addition rule" presented in \eref{theta1.addition.identity}
\begin{equation}
    \vartheta_1(u_1\pm u_2) = \frac{\vartheta_1^2(u_1)\,\vartheta_4^2(u_2)-\vartheta_4^2(u_1)\,\vartheta_1^2(u_2)}{\vartheta_4^2} ~.
\end{equation}
This immediately leads to a significant simplication of $I_{USp(4)}^{(2)}(q,z_{1,2})$, which now becomes
\begin{equation}
    I_{USp(4)}^{(2)}(q,z_{1,2}) = \frac{\eta^4\,\vartheta_3^4}{\vartheta_2^4\,\vartheta_4^4}\left( \frac{\vartheta_1^2(u_1)}{\vartheta_4^2(u_1)} - \frac{\vartheta_1^2(u_2)}{\vartheta_4^2(u_2)} \right)^2 ~.
\end{equation}
Recalling the definition of the Jacobi sine function in \eref{eq:Jacobi_elliptic_def} and \eref{theta.eta}, we recognise this as
\begin{equation}
    I_{USp(4)}^{(2)}(q,z_{1,2}) = 2^{-\frac43}\big[ \lambda(1-\lambda) \big]^{-\frac23}\big( t_1-t_2 \big)^2 ~.
\end{equation}

As expained in and around \eref{eq:Jacobian}, the variable change in \eref{eq:var_change_USp4} leads to the Jacobian
\begin{equation}
    \prod_{j=1}^2 \frac{{\rm d}z_j}{2\pi iz_j} = \frac{1}{4\pi^2\,\vartheta_3^4} \prod_{j=1}^2 \left[ \frac{{\rm d}t_j}{\sqrt{t_j(1-t_j)(\lambda-t_j))}} \right] ~.
\end{equation}
Assembling all the pieces together, the generalised Schur partition function for the $USp(4)$ theory can be expressed as
\begin{equation}
    \cZ_{USp(4)}(q;\alpha) \propto \frac{\eta^4}{\vartheta_3^4}\,\big[ \lambda(1-\lambda) \big]^{-\frac{8\alpha}{3}} \oint_{\mathcal C_2} {\rm d}t_1\,{\rm d}t_2 \prod_{j=1}^2 \bigg[ t_j(1-t_j)(\lambda-t_j) \bigg]^{\alpha-\frac12}\,(t_1-t_2)^{2\alpha} ~.
\end{equation}

The final piece of the proof is the contour $\mathcal C_2$. By repeating the same manipulations performed for the contour $\mathcal C_1$ in the case of the $SU(2)$ generalised Schur partition function, we obtain that both the variables are integrated from $0$ to $\lambda$,
\begin{equation}
    \oint_{\mathcal C_2} {\rm d}t_1\,{\rm d}t_2 = \int_0^\lambda \int_0^\lambda {\rm d}t_1\,{\rm d}t_2 ~.
\end{equation}
Therefore, the generalised Schur partition function of the $USp(4)$ SCFT can finally be expressed as
\begin{equation}
    \cZ_{USp(4)}(q;\alpha) \propto \big[ \lambda(1-\lambda) \big]^{\frac{1-8\alpha}{3}} \oint_0^\lambda {\rm d}t_1 \oint_0^\lambda{\rm d}t_2 \prod_{i=1}^2 \bigg[ t_i(1-t_i)(\lambda-t_i) \bigg]^{\alpha-\frac12}\,(t_1-t_2)^{2\alpha} ~.
\end{equation}
This corresponds to $J_2$ with the parameters
\begin{equation}
    a = \alpha-\frac12 \;,\; b = 2\alpha \;\implies\; p = \frac{1-8\alpha}{3} ~,
\end{equation}
with the appropriate prefactor computed from \eref{prefac}. Thus, we have shown that $\cZ_{USp(4)}(q;\alpha)$ is equal to the contour integral $J_2$ in \eref{JAp} with $n=2$. This solves the third-order MLDE with vanishing Wronskian index \cite{Mathur:1988gt, Naculich:1988xv}:
\be
\left(\cD^3+\mu_2 E_4 \cD+\mu_3 E_6\right)\chi=0
\ee
\eref{MLDEcoeffs.2} then follows by comparing the exponents of the contour integrals to those of the MLDE solutions. This concludes the proof of Proposition \ref{prop:USp2N} for $N=2$. \hfill$\square$

Reading off the CFT data of the solutions from \eref{MLDEcoeffs.2}, we find:
\be
\gamma_0 =-\frac43\alpha +\frac16,\quad \gamma_1=-\frac13\alpha+\frac16,\quad \gamma_2=\frac53\alpha+\frac16
\ee
or equivalently:
\be
c = 32\alpha-4 \;\;,\;\; h_1 = \alpha \;\;,\;\; h_2 = 3\alpha
\label{USp4data}
\ee
The coefficients $\mu_2,\mu_3$ of the MLDE (which are the same as $\nu_2,\nu_3$ since there is a unique modular form of weights 4 and 6)) are then found via \eref{indicial.neq3}:
\be
\mu_2= \frac{1}{36}-\frac{7\alpha^2}{3},\quad \mu_3=\frac{1}{216}(1-8\alpha)(1-2\alpha)(1+10\alpha)
\ee

Note that we have only a one-parameter family of integrals, corresponding to the fact that the generalised Schur partition function of \cite{Deb:2025ypl} has a single free parameter. This means that the family of solutions of the third-order MLDE that we have arrived at is the one-parameter sub-family of the general solution obtained by setting $b=2a+1$.

The case $\alpha=1$, corresponding to the Schur index, results in the character $\chi_{2}$ for the second non-trivial primary (of dimension $h_2=3$) of a $c=28$ vertex algebra whose identity character is logarithmic (this is typical when conformal dimensions differ by an integer, which then lead to a ``resonance'' condition, except in special cases where the resonance obstruction vanishes). An alternative -- and more standard -- interpretation of the Schur index of the $USp(4)$ SCFT is as the identity character of a vertex algebra obtained by permuting the solutions such that the new identity character is the $h_2=3$ character of the original $c=28$ RCFT. From \eref{presexchange}, its new central charge and conformal dimensions $(c,h_1,h_2)$ would be $(c-24h_2, h_1-h_2, -h_2)$ relative to the old values, which works out to $(-44,-2,-3)$. Examining the would-be identity character for such a theory, we find two encouraging features: (i) its $q$-series has positive integral coefficients when normalised to start with 1, which makes it a valid candidate for an identity character, (ii) the second term in the $q$-series is 66 which is the dimension of $SO(12)$, and there is a level $k$, namely $-4$, such that the central charge of $SO(12)$ is $-44$. This suggests that the desired non-unitary vertex algebra has the affine vertex algebra $V^{-4}(SO(12))$ as a vertex operator subalgebra, as expected from the 4d/2d correspondence \cite{Beem:2013sza}.

Another interesting case is the non-unitary minimal model $\cM(2,7)$ with central charge $c=-\frac{68}{7}$ and $(h_1,h_2)=(-\frac27,-\frac37)$. In this form, these numbers do not fit our formula \eref{USp4data}. However one can change the ``presentation'' by treating the third character as the identity as explained around \eref{presexchange} (this procedure is discussed explicitly in \cite{Mukhi:2020gnj} for the three-character case). Then the effective ``central charge'' changes to $c-24h_2=\frac47$. Now the conformal dimensions of the other characters become $\frac17,\frac37$ and we have a perfect match for $\alpha=\frac17$. Note however that this change of presentation, despite making $c$ and all the $h_A$ positive, does not render the theory unitary. The vacuum character of this non-unitary minimal model is the associated VOA of the $(A_1,A_4)$ Argyres--Douglas theory \cite{Cordova:2015nma,Buican:2017uka}.

It is interesting to ask if there are any {\em unitary} three-character VOAs whose non-identity character matches the generalised Schur partition function for some $\alpha$. Remarkably there is just one, which arises at $\alpha=\frac15$ and therefore has $c=\frac{12}{5}$. It is just the $D$-series theory for $SU(2)_8$, discussed long ago in \cite{Cappelli:2009xj}, which has two independent conformal dimensions, $h=\frac15$ and $\frac35$ \footnote{This theory also made an appearance more recently in \cite{Das:2022uoe} and \cite{Rayhaun:2023pgc} in the classification of 3-character/4-primary CFTs.}. For $\alpha=\frac15$ we precisely reproduce the same $c$ and $h_A$, and in the case of three characters and vanishing Wronskian index, this is sufficient to identify all the characters completely. Note that the same set of characters, after choosing the one labelled here as $h=\frac35$ to be the new identity, describes the non-unitary $V_{-\frac85}(SU(2))$ VOA which is known to be associated to the $(A_1,D_5)$ theory.

\subsection{General $N$}
\label{subsec:USp2N_genSchur}

We now consider the general case of the generalised Schur partition function for the $USp(2N)$ SCFT with $2N+2$ fundamental hypermultiplets. In particular, we will argue that 
$\cZ_{USp(2N)}(q;\alpha)$ can be identified with the contour integral $J_N$ in \eref{JAp}, which solves an order $N+1$ MLDE with vanishing Wronskian index. The proof is a simple generalisation of the $N=2$ case as we demonstrate next.

\noindent{\bf Proof} :

From \eref{eq:USpN_genSchur_1par}, the relevant integrand can be expressed as
\begin{equation}
    I_{USp(2N)}(q,\mbf z) = \eta^{2N(N+2)} \prod_{i=1}^N \frac{\vartheta_1^2(2u_i)}{\vartheta_4^8(u_i)} \prod_{1\leq i<j\leq N} \frac{\vartheta_1^2(u_i\pm u_j)}{\vartheta_4^4(u_i)\vartheta_4^4(2u_j)} ~.
\end{equation}
As before, we can split this into the following two pieces:
\begin{equation}
    \begin{split}
        I_{USp(2N)}^{(1)}(q,\mbf z) &= \eta^{6N} \prod_{i=1}^N \frac{\vartheta_1^2(2u_i)}{\vartheta_4^8(u_i)} ~, \\
        I_{USp(2N)}^{(2)}(q,\mbf z) &= \eta^{2N(N-1)} \prod_{1\leq i<j\leq N} \frac{\vartheta_1^2(u_i\pm u_j)}{\vartheta_4^4(u_i)\vartheta_4^4(2u_j)} ~.
    \end{split}
\end{equation}
We notice that these are exactly in the same form as the two pieces in the $USp(4)$ case.

Therefore, by analogy with \eref{eq:var_change_USp4}, we define the variables
\begin{equation}\label{eq:var_change_USpN}
    t_i = \lambda\,{\rm sn}^2(v_i,k) \;,\; \text{ where } v_i = \pi\,\vartheta_3^2\,u_i \;,\; \text{ for } i=1,2,\dots,N ~,
\end{equation}
By using the standard identities discussed before and presented in Appendix \ref{app:special_functions}, the first piece of the integrand can be represented as
\begin{equation}
    I_{USp(2N)}^{(1)}(q,\mbf z) = \prod_{i=1}^N \left[ \frac{t_i(1-t_i)(\lambda-t_i)}{\lambda(1-\lambda)} \right] ~.
\end{equation}
Next, we can analyse the second piece by using the theta addition rule \eref{theta1.addition.identity} to obtain
\begin{equation}
    I_{USp(2N)}^{(2)}(q,\mbf z) = \prod_{1\leq i<j\leq N} \frac{\eta^4\,\vartheta_3^4}{\vartheta_2^4\,\vartheta_4^4}\left( \frac{\vartheta_1^2(u_i)}{\vartheta_4^2(u_i)} - \frac{\vartheta_1^2(u_j)}{\vartheta_4^2(u_j)} \right)^2 ~.
\end{equation}
As before, using \eref{theta.eta} and the definition of $\lambda$, this can be simplified to
\begin{equation}
    I_{USp(2N)}^{(2)}(q,\mbf z) = \prod_{1\leq i<j\leq N} 2^{-\frac43}\big[ \lambda(1-\lambda) \big]^{-\frac23}\big( t_i-t_j \big)^2 ~.
\end{equation}

Under the variable transformation \eref{eq:var_change_USpN} the integrals picks up the usual Jacobian factor
\begin{equation}\label{eq:USp2N_Jacobian}
    \prod_{j=1}^N \frac{{\rm d}z_j}{2\pi iz_j} = \frac{1}{(2\pi)^N\,\vartheta_3^{2N}} \prod_{j=1}^N \left[ \frac{{\rm d}t_j}{\sqrt{t_j(1-t_j)(\lambda-t_j))}} \right] ~.
\end{equation}
Assembling these pieces together, the generalised Schur partition function for the $USp(2N)$ SCFT can be expressed as
\begin{equation}
    \begin{split}
        \cZ_{USp(2N)}(q;\alpha) \propto \frac{\eta^{2N}}{\vartheta_3^{2N}}\,\big[ \lambda(1-\lambda) \big]^{-\frac{N(N+2)}{3}\,\alpha} \oint_{\mathcal C_N} \prod_{j=1}^N {\rm d}t_j &\prod_{i=1}^N \bigg[ t_i(1-t_i)(\lambda-t_i) \bigg]^{\alpha-\frac12} \\
        &\times \prod_{1\leq i<j\leq N}\big( t_i-t_j \big)^{2\alpha} ~.
    \end{split}
\end{equation}
Using the same argument for each variable $t_i$ as discussed in the proof for $\cZ_{SU(2)}(q;\alpha)$, we get the contour
\begin{equation}
    \oint_{\mathcal C_N} = \underbrace{\int_0^\lambda\int_0^\lambda\dots\int_0^\lambda}_{N \;{\rm times}} ~.
\end{equation}
Therefore, the generalised Schur partition function of the $USp(2N)$ SCFT with $2N+2$ fundamental hypermultiplets can finally be expressed as
\begin{equation}\label{eq:USp2N_contour}
    \begin{split}
        \cZ_{USp(2N)}(q;\alpha) \propto \big[ \lambda(1-\lambda) \big]^{\frac{N}{6}\big(1-2\alpha(N+2)\big)} \oint_{\mathcal C_N} \prod_{j=1}^N {\rm d}t_j &\prod_{i=1}^N \bigg[ t_i(1-t_i)(\lambda-t_i) \bigg]^{\alpha-\frac12} \\
        &\times \prod_{1\leq i<j\leq N}\big( t_i-t_j \big)^{2\alpha} ~.
    \end{split}
\end{equation}
This corresponds to $J_N$ in \eref{JAp} with parameters
\begin{equation}
    a = \alpha-\frac12 \;,\; b = 2\alpha \;\implies\; p = \frac{N}{6}\big(1-2\alpha(N+2)\big) ~,
\end{equation}
which agrees with the exponent of $\lambda$ in \eref{eq:USp2N_contour}. Thus, we have shown that $\cZ_{USp(2N)}(q;\alpha)$ is identified with the contour integral $J_A$ and thus solves the order $N+1$ MLDE with vanishing Wronskian index.

Since the generalised Schur partition function depends on a single parameter $\alpha$, it is clear that we can only get a one-parameter sub-family of the original contour integrals \eref{JAp}, and we now see that the required specialisation is $\beta=\alpha-\half$ and $\gamma=2\alpha$. Let us now try to understand which 2d VOA characters can arise in this way. In \eref{chident} we have written out the relation between $c,h_A$ of a putative RCFT and the parameters $a,b$ of a contour integral, and specialising this to our case gives:
\be
\begin{split}
c &= 2N\Big( 2(N+2)\alpha - 1\Big) ~, \\[2mm]
h_A &= \frac{A(A+1)}{2}\alpha \;,\; \text{ where } A=1,\dots,N ~.
\end{split}
\label{chidentUSp}
\ee
Recall that in \eref{chident} $n$ is the number of characters, while $n-1$ is the number of integration variables in the contour integral. In the present case, we have $N$ integration variables (coming from $USp(2N))$, so we must identify $n=N+1$ to obtain the above parametrisation in \eref{chidentUSp}.

It remains to determine the MLDE completely. As indicated below the statement of proposition \ref{prop:USp2N}, for an MLDE of sixth or higher order one can only determine the $\nu_k=\sum_i \mu_{k,i}$ in terms of the exponents in \eref{chidentUSp}, but not the individual $\mu_{k,i}$. One has to expand the contour integral \eref{eq:USp2N_contour} in a $q$-series \footnote{In practice, we must first expand the integral in a series in $\lambda$ to finite order, then evaluate the integrals for each coefficient and finally replace $\la$ by a series in $\sqrt{q}$ -- in which eventually only integral powers of $q$ will contribute.}. Fortunately in the examples below we will be able to rely on an old result of \cite{Mukhi:1989qk} where this step was effectively carried out in many cases. This proves Proposition \ref{prop:USp2N} for general $N$.  \hfill$\square$

We now turn to some examples. First consider $\alpha=\half$. From \eref{chidentUSp} we get:
\be
\begin{split}
c &= 2N(N+1)\\
h_A &=\frac{A(A+1)}{4},\quad A=1,2,\cdots,N
\end{split}
\ee
If now we choose the highest-dimension primary to be the new vacuum character, this replaces $c$ by $c-24h_N$ which gives:
\be
\begin{split}
c &= -4N(N+1)
\end{split}
\ee
We recognise this as the central charge of $V_{-N-\half}\big(SU(2N+1)\big)$. This case has been associated in \cite{Deb:2025ypl} with the Argyres--Douglas theory $D_2(SU(2N+1))$ whose associated VOA is given by this simple affine vertex algebra \cite{Xie:2016evu,Beem:2017ooy}.

For the next example, consider the Schur index of the $USp(2N)$ theory which corresponds to setting $\alpha=1$ in \eref{chidentUSp}. This gives:
\be
\begin{split}
c &= 2N(2N+3),\\
h_A &=\frac{A(A+1)}{2} ~.
\end{split}
\ee
Thus all the conformal dimensions are integral, and typically this means the set of characters has at least one with logarithmic behaviour in $q$. To match this series with the contour integrals \eref{eq:USp2N_contour} we again switch presentation by choosing the ``last'' primary to be the new identity. Then we get:
\be
c\to c-24h_N=-2N(4N+3)
\ee
which agrees with $-12c_{4d}$ as quoted in \cite{Deb:2025ypl}.

Now consider the non-unitary $(2,2r+1)$ minimal models with $r=2,3,\cdots$. In the standard BPZ form they have:
\be
\begin{split}
c &=-\frac{2(r-1)(6r-1)}{2r+1} ~, \\[2mm]
h_A &=-\frac{A(2r-1-A)}{2(2r+1)},\quad A=1,2,\dots, r-1 ~.
\end{split}
\ee
This theory has $r$ independent characters. Let us re-organise these fields by treating the primary of highest $|h|$ as the identity character. From the above formula, this primary has dimension $h_{r-1}$. Hence when we do this, the central charge shifts to $c-24h_{r-1}$ while $r-2$ conformal dimensions shift to $h_A-h_{r-1}, A=1,2,\dots, r-2$ while the last one (corresponding to the original identity character) becomes $-h_{r-1}$. In this way, both $c$ and all the $h_A$ become positive \footnote{As already mentioned in the $N=2$ case, this does not mean the theory is unitary.}. In particular the central charge becomes:
\be
c = \frac{2(r-1)}{2r+1} ~,
\ee
This matches perfectly with \eref{chidentUSp} if we identify $n=r$ and choose $\alpha=\frac{1}{2r+1}$. Now we consider the conformal dimensions, and relabel $A\to r-1-A$ to find:
\be
h_A=\frac{A(A+1)}{2(2r+1)} ~,
\ee
which also agrees with \eref{chidentUSp}. Recalling that $n=N+1$, we see that for the $USp(2N)$ 4d theory there is a matching of exponents between the generalised Schur partition function with $\alpha=\frac{1}{2N+3}$ and the identity character of the $(2,2N+3)$ non-unitary minimal model, considered in a special presentation where the primary of most negative weight is taken as the identity. Recall that this corresponds to the associated VOA of the $(A_1,A_{2N})$ Argyres--Douglas theory, and hence the last character in the ``unitary'' presentation, or equivalently the identity character of the original non-unitary presentation, is identified with the Schur index of this non-Lagrangian SCFT \cite{Beem:2017ooy}. 

Finally we ask whether there are any unitary RCFTs for which the generalised Schur partition function, written as in \eref{eq:USp2N_contour}, is the character of the ``last primary''. We already saw an example, namely the $\cE_3(SU(2)_8)$ example for $USp(4)$ that was discussed earlier. This theory is the $D$-series invariant of $SU(2)_8$, and it is known that such invariants arise at all levels $k$ that are a multiple of 4. Moreover the conformal dimensions of the primaries are given by $\frac{j(j+1)}{k+2}$ where $j=1,2,\cdots \frac{k}{4}$ and the number of characters is $\frac{k}{4}+1$. This leads to a series of unitary CFTs of the form $\cE_{N+2}(SU(2)_{4N})$ whose central charge and conformal dimension are:
Their central charge is:
\be
\begin{split}
c &=\frac{6N}{2N+1}\\[2mm]
h_A &= \frac{A(A+1)}{2(2N+1)}, \quad A=1,2,\cdots, N
\end{split}
\ee
It is easy to verify that this conformal data perfectly matches that in \eref{chidentUSp} for the generalised Schur index of the 4d $USp(2N)$ theory, with $\alpha=\frac{1}{2N+1}$. Thus (subject to the caveat in the following paragraph) the generalised Schur index for this $\alpha$ is the last character of the above theory. As discussed previously, it can also be associated with the first character of the non-unitary theory obtained by sending $c\to c-24h$. Doing this, we find $c=-6N$. This agrees with $-12c_{4d}$, and this family has been identified  with the $(A_1,D_{2N+1})$ Argyres-Douglas theories. 

In this example, only for $N=2,3,4$ is the precise matching of indices sufficient to guarantee agreement of the generalised Schur index of $USp(2N)$ with the RCFT character. For $N\ge 5$ we have $\ge 6$ characters and, as mentioned before, in such cases matching of indices is not sufficient to identify characters. Fortunately this family of $D$-series $SU(2)_k$ theories is among those for which \cite{Mukhi:1989qk} had long ago conjectured complete equivalence of the characters with contour integrals, and provided some non-trivial evidence for this claim.

Thus we have reproduced all four cases in Table III of \cite{Deb:2025ypl} corresponding to the values $\alpha=\half,1,\frac{1}{2N+3},\frac{1}{2N+1}$ of the parameter of the generalised Schur partition function.

\subsection{Two-parameter generalised Schur partition function}
\label{subsec:USp2N_genSchur_2par}

In this section, we propose a natural two-parameter extension of the generalised Schur partition function introduced in \cite{Deb:2025ypl}. While we are not able to obtain this extension from a particular limit of the full superconformal index, we nevertheless demonstrate that this {\it two-parameter generalised Schur partition function} satisfies an order $N+1$ MLDE with vanishing Wronskian index.

The inspiration for this generalisation follows from the proof of Proposition \ref{prop:USp2N}. Recall that we proved that $\cZ_{USp(2N)}(q;\alpha)$ yields a one-parameter sub-family of the original contour integrals reviewed in \eref{JAp}. It is then natural to ask if there is an extension of $\cZ_{USp(2N)}(q;\alpha)$ that is equivalent to the full two-parameter family of the contour integrals \eref{JAp}.

To this end, we propose the following as the two-parameter generalised Schur partition function associated to the $USp(2N)$ SCFT with $2N+2$ fundamental hypermultiplets:
\begin{equation}\label{eq:USp2N_genSchur_2par}
    \begin{split}
        &\cZ_{USp(2N)}(q;\alpha,\beta) = \\
        &\quad \;\frac{\eta^{2N}}{|W|\,N(\alpha)}\oint \prod_{j=1}^N \frac{{\rm d}z_j}{2\pi iz_j} \prod_{i=1}^N \left( \eta^6\,\frac{\vartheta_1^2(2u_i)}{\vartheta_4^8(u_i)} \right)^\alpha \prod_{1\leq i <j\leq N} \left( \eta^4\,\frac{\vartheta_1^2(u_i\pm u_j)}{\vartheta_4^4(u_i)\vartheta_4^4(u_j)} \right)^\beta ~.
    \end{split}
\end{equation}
It is immediately clear that the $\alpha=\beta$ case of $\cZ_{USp(2N)}(q;\alpha,\beta)$ is equivalent to the generalised Schur partition function, $\cZ_{USp(2N)}(q;\alpha)$, introduced by \cite{Deb:2025ypl}. Therefore for $\alpha=\beta=1$, this furnishes the Schur index of the $USp(2N)$ SCFT, and thus $\cZ_{USp(2N)}(q;\alpha,\beta)$ can be viewed as a two-parameter generalisation of the Schur index for this gauge theory.

As is natural from our motivation, we make the following proposition:
\begin{prop}\label{prop:USp2N_2par}
    The two-parameter generalised Schur partition function $\cZ_{USp(2N)}(q;\alpha,\beta)$, with analytically continued $\alpha,\beta\in\mathbb C$, is a solution to the following order-$(N+1)$ modular linear differential equation with vanishing Wronskian index: 
    \begin{equation}\label{eq:USp2N_MLDE_2par}
        \left(\mathcal D^{N+1} + \sum_{k=1}^{N+1} \sum_i \mu_{k,i} \, \phi_{2k,i}(\tau)\, \mathcal D^{N+1-k}\right)\,\cZ_{USp(2N)}(q;\alpha,\beta) = 0 ~,
    \end{equation}
    where $\sum_i \mu_{k,i}$ is determined by \eref{indicial} as before, but with:
    \begin{equation}\label{MLDEcoeffs.4}
        \begin{split}
            \gamma_0 &=-\frac{N}{12}\Big(6\alpha+2(N-1)\beta-1\Big) ~, \\[2mm]
            \gamma_A &= \gamma_0+\alpha A+\frac{\beta}{2}A(A-1) ~.
        \end{split}
    \end{equation}
\end{prop}
Again, it must be emphasised that with  additional work it is straightforward to determine all the $\mu_{k,i}$, and hence the MLDE, completely by expanding $\cZ_{USp(2N)}$ in a $q$-series.

\noindent{\bf Proof} : 

This proof is almost identical to the one considered in Section \ref{subsec:USp2N_genSchur} and hence we do not provide many details here. The integrand of $\cZ_{USp(2N)}(q;\alpha,\beta)$ naturally has two components that can be expressed as
\begin{equation}
    \begin{split}
        I_{USp(2N)}^{(1)}(q,\mbf z) &= \prod_{i=1}^N \eta^6\,\frac{\vartheta_1^2(2u_i)}{\vartheta_4^8(u_i)} \\
        I_{USp(2N)}^{(2)}(q,\mbf z) &= \prod_{1\leq i <j\leq N} \eta^4\,\frac{\vartheta_1^2(u_i\pm u_j)}{\vartheta_4^4(u_i)\vartheta_4^4(u_j)}
    \end{split}
\end{equation}
Thus by identical arguments as in the proof of Proposition \ref{prop:USp2N}, after the same variable redefinition
\begin{equation}\label{eq:var_change_USpN_2par}
    t_i = \lambda\,{\rm sn}^2(v_i,k) \;,\; \text{ where } v_i = \pi\,\vartheta_3^2\,u_i \;,\; \text{ for } i=1,2,\dots,N ~,
\end{equation}
these two pieces in the integrand become
\begin{equation}
    \begin{split}
        I_{USp(2N)}^{(1)}(q,\mbf z) &= \prod_{i=1}^N \left[ \frac{t_i(1-t_i)(\lambda-t_i)}{\lambda(1-\lambda)} \right] \\
        I_{USp(2N)}^{(2)}(q,\mbf z) &= \prod_{1\leq i<j\leq N} 2^{-\frac43}\big[ \lambda(1-\lambda) \big]^{-\frac23}\big( t_i-t_j \big)^2 ~.
    \end{split}
\end{equation}
The Jacobian from this variable change is the same as in \eref{eq:USp2N_Jacobian}. Finally, the corresponding contour is identical as well due to identical variable transformations. Therefore, the final expression for the two-parameter generalised Schur partition function $\cZ_{USp(2N)}(q;\alpha,\beta)$ can be expressed as
\begin{equation}\label{eq:USp2N_contour_2par}
    \begin{split}
        \cZ_{USp(2N)}(q;\alpha) \propto \big[ \lambda(1-\lambda) \big]^{\frac{N}{6}\big(1-6\alpha+2\beta-2N\beta\big)} \oint_{\mathcal C_N} \prod_{j=1}^N {\rm d}t_j &\prod_{i=1}^N \bigg[ t_i(1-t_i)(\lambda-t_i) \bigg]^{\alpha-\frac12} \\
        &\times \prod_{1\leq i<j\leq N}\big( t_i-t_j \big)^{2\beta} ~.
    \end{split}
\end{equation}
This corresponds to $J_N$ in \eref{JAp} with parameters
\begin{equation}
    a = \alpha-\frac12 \;,\; b = 2\beta \;\implies\; p = \frac{N}{6}\big( 1-6\alpha+2\beta-2N\beta \big) ~,
    \label{eq:USp_2par_ident}
\end{equation}
which agrees with the exponent of $\lambda$ in \eref{eq:USp2N_contour_2par}. Thus, as expected, the two-parameter generalised Schur partition function $\cZ_{USp(2N)}(q;\alpha,\beta)$ can be identified with the contour integral $J_A$ and thus solves the order $N+1$ MLDE with vanishing Wronskian index. The identification of MLDE parameters $\nu_k=\sum_i\mu_{k,i}$ with the critical exponents $\gamma_A$ follows as before by comparing the leading behaviour of contour integrals with that of MLDE solutions, and the entire set $\mu_{k,i}$ can then be determined by expanding the solution in a $q$-series. This then proves Proposition \ref{prop:USp2N_2par}.  \hfill$\square$

For $N=2$, this implies in particular that the MLDE coming from the two-parameter generalised Schur index is solved by all three-character RCFT with $\ell=0$. Combining Eqs.(\ref{chident}) and (\ref{eq:USp_2par_ident}), we get:
\be
\begin{split}
c &=8\left(3\alpha+\beta-\frac12\right)\\
h_1 &= a+\half=\alpha\\
h_2 &= 2\alpha+\beta
\end{split}
\ee
which gives:
\be
\alpha= h_1, \qquad \beta=\frac{c}{8}+\half -3h_1
\ee
Note that $h_2$ is then determined by the Riemann-Roch identity \eref{RiemRoch}. For a unitary theory, it is natural to identify the three contour integrals consecutively with the exponents in the order $\gamma_0 < \gamma_1 <\gamma_2$ but in general there is a freedom in the above identification since we can match any of the solutions with any of the contour integrals.

The set of $\ell=0$ RCFTs with three characters includes the Ising model, all $SO(N)_1$ WZW models \cite{Mathur:1988gt}, the coset models in \cite{Gaberdiel:2016zke} and several additional examples in \cite{Mukhi:2020gnj, Kaidi:2021ent, Das:2021uvd, Bae:2021mej, Duan:2022kxr, Rayhaun:2023pgc}. The complete classification (excluding models with $c=8,16$ which have special features) was presented in \cite{Das:2022uoe}. A few illustrative examples are listed in Table \ref{Table:threechar}.

Returning to the case of general $N$, now that we have two parameters we can simply take over all the results of \cite{Mukhi:1989qk} where characters of known RCFT are matched with contour integral representations. The list of such CFTs in this reference includes the $A$, $D$ and $E$-series characters based on $SU(2)_k$ RCFTs, as well as the $SU(N)_1$ characters. A detailed list of WZW models with $\ell=0$ is provided in \cite{Das:2020wsi} so one can also try to match additional Kac-Moody algebras besides those of $SU(N)$. Here it becomes essential to work out a general proof that entire characters are matched and not just critical exponents. We hope to report on these issues in the future \cite{Chandra:2026inprogress}. 

\begin{table}[h!]
\centering
\begin{tabular}{|c|c|c|c|}
\hline
\Tstrut $(c\,|\,h_1,h_2)$ & $\alpha$ & $\beta$ & Description \\[2mm]
\hline
\Tstrut $\left(\half\,|\,\frac{1}{16},\half\right)$ & $\frac{1}{16}$ & $\frac{3}{8}$ & Ising model \\[2mm]
$\left(4\,|\,\frac25,\frac35\right)$ & $\frac25$ & $-\frac15$ & $SU(5)_1$ WZW Model\\[2mm]
$\left(\frac{52}{5}\,|\,\frac35,\frac65\right)$ & $\frac35$ & 0 & $F_{2,1}\otimes F_{2,1}$ WZW model\\[2mm]
$\left(\frac{68}{5}\,|\,\frac45,\frac75\right) $ & $\frac45$ & $-\frac15$ & ${\cal E}_3\big(USp(16)_1\big)$\\[2mm]
$\left(15\,|\,\frac78,\frac32\right)$ & $\frac78$ & $-\frac14$ & ${\cal E}_3\big(SU(16)_1\big)$\\[2mm]
$\left(\frac{31}{2}\,|\,\frac{15}{16},\frac32\right)$ & $\frac{15}{16}$ & $-\frac38$ & $E_{8,2}$ WZW model\\[2mm]
$\left(\frac{92}{5}\,|\,\frac65,\frac85\right)$ & $\frac65$ & $-\frac45$ & $\cE_3\big(E_{6,3}\,G_{2,1}\big)$\\[2mm]
$\left(20\,|\,\frac43,\frac53\right)$ & $\frac43$ & $-1$ & $\cE_3\big(SU(3)_1)^{10}\big)$\\[2mm]
$\left(\frac{47}{2}\,|\,\frac32,\frac{31}{16}\right)$ & $\frac32$ & $-\frac{17}{16}$ & Baby Monster CFT\\[2mm]
$\left(\frac{n}{2}\,|\,\frac12,\frac{n}{16}\right)$ & $\half$ & $\frac{n}{16}-1$ & $SO(n)_1$ WZW models\\[2mm]
\hline
\end{tabular}
\caption{Illustrative list of three-character RCFTs having a character that matches the two-parameter generalised Schur partition function of $USp(4)$. Here $\cE_3$ denotes a three-character extension of the given Kac-Moody algebra.}
\label{Table:threechar}
\end{table}

\section{Discussion and open questions}
\label{sec:Discussiob}

In this section, we summarise the concrete calculations of this work, formulate a conjecture, and outline possible generalisations. While it also serves as a conclusion, it is more than just a summary.

In this work, we have demonstrated a structural relationship between the generalised Schur partition function $\cZ_G(q;\alpha)$ introduced in \cite{Deb:2025ypl} and a multi-variable contour integral representation of vector-valued modular forms that was proposed in \cite{Mukhi:1989qk}. Concretely, we proved that the analytically continued $\cZ_G(q;\alpha)$ of the 4d $\cN=2$ $USp(2N)$ SCFT with $2N+2$ fundamental hypermultiplets (in particular, the $N=1$ case: $SU(2)$ SCFT with $N_f=4$) satisfies an order $N+1$ modular linear differential equation with vanishing Wronksian index. Along the way, we proposed a two-parameter generalisation of the Schur index of the $USp(2N)$ SCFT that naturally extends the one-parameter case introduced in \cite{Deb:2025ypl}.

A direct byproduct of this work is the algorithmic determination of the modular data associated to the Schur indices of 4d SCFTs that appear at special values of $\alpha$.\footnote{See \cite{Arakawa:2025dcl} for some recent work on the determination of modular data for boundary admissible affine $\cW$-algebras. These appear as the associated VOA of some of the Argyres--Douglas theories obtained from $\cZ_{USp(2N)}(q;\alpha)$.} While the modular $\cT$-matrix is simply encoded in the indicial roots of the MLDE, the full modular $\mathcal S$-matrix can be determined by using the algorithm proposed in \cite{Mukhi:2019cpu} that utilises these contour integral representations. This serves as useful input data for analyses along the lines of \cite{ArabiArdehali:2023bpq,Pan:2024bne}.

We note here that the appearance of characters of unitary RCFTs in $\cZ_{USp(2N)}(q;\alpha)$ for special values of $\alpha$, as discussed in the main body of this work, is intriguing and has already been observed in \cite{Buican:2017rya}. Although we demonstrate this by showing that they arise as solutions to the same one-parameter sub-family of an MLDE, a direct 4d interpretation of these SCFT/unitary RCFT connections remains to be understood.\footnote{A promising direction, based on the discussion in Subsection \ref{subsec:monodromy} might be to consider twisted circle compactifications based on and explored in the following series of works \cite{Fredrickson:2017yka,Dedushenko:2018bpp,ArabiArdehali:2024vli,ArabiArdehali:2024ysy,Gang:2018huc,Gang:2021hrd,Gang:2023rei,Gang:2024loa}.}

We now proceed to examine the broader implications of our results, beginning with the relation to monodromy traces.

\subsection{Relation to BPS monodromy traces and MLDE}
\label{subsec:monodromy}

The ($q$-deformed) Kontsevich--Soibelmann monodromy operator $M$, simply referred to as the quantum/BPS monodromy operator in this context, is a wall-crossing invariant (up to conjugations) that encodes the BPS spectrum on the Coulomb branch \cite{Cecotti:2010fi}.

In this section, we discuss the relation between ${\rm Tr}\,M^k$ and the one-parameter generalised Schur partition function $\cZ_G(q;\alpha)$, along with the implications that follow. We first note a parametric connection between the $\cZ_G(q;\alpha)$ and the specialised index introduced in \cite{Cecotti:2015lab}. Both of these are obtained from a similar limit, and the specialised index in turn is related to the trace of powers of the BPS monodromy operator. We point out some of the subtleties that are involved. The various traces of the BPS monodromy operator, ${\rm Tr}\, \cM^k$,\footnote{The two conventions of monodromy traces $\cM$ and $M$ in \cite{Cecotti:2015lab} are related by an overall ``massless photon" factor: $\cM = (q;q)_\infty^{-2r}\,M$.}  are associated with vertex operator algebras $\cV^{(k)}$ whose central charges satisfy a given conjectural relation. We demonstrate and prove this relation directly for the central charges obtained from $\cZ_{USp(2N)}(q;\alpha)$. We will end the section with a conjecture motivated by these connections and empirical observations.

The specialised index was defined in \cite{Cecotti:2015lab} to be the following limit of the superconformal index:
\begin{equation}
    {\rm Tr}\,\cM^k \; \sim \; \cI_k^{\rm spec}(q) = \cI(p,q,t)_{p\rightarrow e^{2\pi i},\,t=q\,p^{k+1}}  \;,\; \text{ for } k\in\mathbb Z ~,
\end{equation}
where the equality between ${\rm Tr}\,\cM^k$ and $\cI_k^{\rm spec}(q)$ in full generality is conjectural and subject to regularisation, as we will briefly discuss below. Considering $t=q\,p^{k+1}$ in isolation, this limit can be simplified into
\begin{equation}
    \log t = \left( 1+k\frac{\log p}{\log{q\,p}} \right)\log{q\,p} = \left( 1-k\frac{\epsilon}{\log q} \right) \log{q\,p} ~,
\end{equation}
where we have introduced a new parameter $\epsilon=- \frac{\log q\log p}{\log{q\,p}}$, that implies that
\begin{equation}
    \log p = -\frac{\epsilon}{1+\frac{\epsilon}{\log q}} \;\implies\; p = 1 - \epsilon + \mathcal O(\epsilon^2) ~,
\end{equation}
in the limit $\epsilon\rightarrow0$. This reproduces the same limit as the generalised Schur limit in \eref{eq:gen_Schur_limit} under the identification
\begin{equation}
    k = -\alpha \,\in\, \mathbb Z ~.
\end{equation}

However, this identification between the generalised Schur partiton function (for integral $\alpha$) with the specialised index is subtle. This is mostly due to certain zero modes of the vector multiplet. For the generalised Schur partition function, this is best seen from the single letter index of the vector multiplet in \eref{eq:singleletter_genSchur},
\begin{equation}
    i_V(q,\mbf z;\alpha) = {\rm PE}\big[(1-\alpha)\,{\rm rk}\,G\big] \times \Delta(\mbf z)^{\alpha-1} \, {\rm PE\left[ -\frac{2q}{1-q}\chi_{\rm adj}(\mbf z) \right]} ~,
\end{equation}
where the ${\rm PE}\big[(\alpha-1)\,{\rm rk}\,G\big]$ diverges for $\alpha<1$ and goes to zero for $\alpha>1$. As discussed in Section \ref{subsec:generalised_schur}, this can be regulated in a particular way that leads to the $(q;q)^{2\,{\rm rk}\,G}$ factor in \eref{eq:NoNorm_GenSchur_def}. On the other hand, it was observed that there is an analogous divergence/zero in the limit $\cI_k(q)$ associated to the vector multiplet for $k>0$ or $k<0$, respectively. For the detailed characterisation, we refer to \cite{Cecotti:2015lab}, where it was also shown that this can be regularised by inserting a line defect in a way that exactly cancels these zero modes.

To match the generalised Schur partition function and the specialised index, one must then ask if these two regularisation schemes coincide. A simple observation for the $SU(2)$ SCFT with $4$ fundamental hypermultiplets reveals a disagreement. The difference being that the Vandermonde determinant, $\Delta(\alpha)$, is raised to power $\alpha$ in $\cZ_{SU(2)}(q;\alpha)$, whereas the $k=-\alpha=-2$ example presented in \cite{Cecotti:2015lab}, for instance, has exactly $\Delta(\alpha)$ in the integration measure. Therefore, while we believe that there is a direct relation between $\cZ_G(q;-k)$ and $\cI_k^{\rm spec}(q) \sim {\rm Tr}\cM^k$, there are some subtleties that we do not explore further in this work.\footnote{For instance, the insertion of line defects in the infrared formulas for the trace of powers of BPS monodromy operators can produce the appropriate power of the Vandermonde determinant.}

Turning to the purely infrared computation of ${\rm Tr}\,\cM^k$, we consider the empirical observations made recently in \cite{Kim:2024dxu}. These authors showed that a series of traces of powers of the BPS monodromy operators for some low-rank Argyres--Douglas theories are identified with characters of unitary 2d RCFT characters; many of these were part of the MMS series of unitary 2d RCFTs. These results are in agreement with the analysis of ${\rm Tr}\,\cM$ of the families Argyres--Douglas theories considered in \cite{Cecotti:2015lab}. Futhermore, in \cite{Cecotti:2015lab}, it was shown that ${\rm Tr}\,\cM$ for the $(A_1,D_{2N+1})$ Argyres--Douglas theory is related to the vacuum character of the $SU(2N+1)_1$ WZW model. Recall that the Schur index of this theory satisfies an order-($N+1$) $\ell=0$ MLDE; the same is true for characters of the $SU(2N+1)_1$ WZW model.

Building on the results of \cite{Cecotti:2015lab}, it is proposed in \cite{Kim:2024dxu} that the trace of a higher power of the monodromy matrix
\begin{equation}
    I_k(q) = (q;q)_\infty^{2r}\,{\rm Tr}\,M^k ~,
\end{equation}
when convergent, produces the vacuum character of a vertex algebra, $\cV^{(k)}$, with leading $q$ power 1. The central charges of these vertex algebras was proposed to be
\begin{equation}\label{eq:KimSong_ccRelation}
    c_{\rm 2d}^{(k)} = 12k\,c_{\rm 4d} - 2r(k+1) ~.
\end{equation}
In this expression, $c_{\rm 4d}$ is the 4d central charge of the original 4d $\cN=2$ SCFT. Note that in this language, $\cV^{(-1)}$ corresponds to the standard associated non-unitary VOA of the 4d $\cN=2$ SCFT \cite{Beem:2013sza} with the standard relation between the 4d and 2d central charges.

Here, we observe that $c_{2d}^{(k)}$ for the case of the $USp(2N)$ SCFT is simply the 2d central charge associated to the non-unitary presentation of the generalised Schur partition function for $\alpha=-k$. Concretely, recall the exponents of the putative RCFT corresponding to $\cZ_{USp(2N)}(q;\alpha)$ presented in \eref{chidentUSp}. Consider the ``effective" central charge, $c_{\rm eff}(\alpha)$, associated to the presentation with respect to the $h_N$ conformal dimension (recall that this corresponds to the contour integral \eref{eq:USp2N_contour}):
\begin{equation}
    \begin{split}
        -\frac{c_{\rm eff}(\alpha)}{24} = h_N - \frac{c}{24} &= \frac{N(N+1)}{2}\,\alpha - \frac{N\big(2(N+2)\,\alpha-1\big)}{12} \\
        &= \frac{N\big(2(2N+1)\,\alpha+1\big)}{12} ~.
    \end{split}
\end{equation}
Using this, it can be immediately seen that $c_{\rm eff}(-k)$ satisfies \eref{eq:KimSong_ccRelation} for
\begin{equation}
    c_{\rm 4d} = \frac{N(4N+3)}{6} \;,\quad r = N ~,
\end{equation}
where $c_{\rm 4d}$ is the central charge of the $USp(2N)$ SCFT with $2N+2$ fundamental hypermultiplets, and $r=N$ is the rank of this SCFT, {\it i.e.}, the complex dimension of its Coulomb branch of the moduli space of vacua.

As an aside, we note that \eref{eq:KimSong_ccRelation} is exactly the necessary condition for the central charges that was described in \cite{Deb:2025ypl} for the partition function of two theories to be related to each other. Here we provided an explicit derivation of this relation for the $USp(2N)$ SCFTs that we consider. We believe that the above necessary condition, for general $G$, originates from the relation between the generalised Schur partition function and its associated MLDE. Note that \cite{Deb:2025ypl} conjectures an additional necessary condition that relates the Coulomb branch scaling dimensions of theories connected by the same generalised Schur partition function. We suspect that such a condition can also be proved by relating these scaling dimensions to other indicial roots of the MLDE. We leave a proper treatment for future work.

Motivated by these observations and our primary result that the generalised Schur partition function satisfies a fixed order and $\ell$ MLDE, we make the following conjecture:
\begin{conj}\label{conj1}
    For any given 4d $\cN=2$ superconformal field theory, the following trace of $k$-th powers of the associated BPS monodromy matrix:
    \begin{equation}
        \cI_k(q) \coloneq q^{\frac12 c(k)}(q;q)_\infty^{2r}\,{\rm Tr}\, M^k ~,
    \end{equation}
    solves a modular linear differential equation of fixed order and Wronksian index, for all real (and also complex) $k$. Here, $r$ is the rank of the SCFT and
    \begin{equation}
        c(k) = -k\,c_{\rm4d}+\frac{r\,(k+1)}{6} ~,
    \end{equation}
    where $c_{\rm 4d}$ is the 4d central charge.
\end{conj}

It is important to remember here that for large enough integers $|k|$, ${\rm Tr}\,M^k$ diverges and this conjecture is subject to an appropriate regularisation, particularly so in the case of non-integral $k$. For $k=-1$, this reduces to the statement that the Schur index, which itself is conjecturally equal to $\cI_{-1}(q)$, solves an MLDE -- this has been checked extensively. Thus according to Conjecture \ref{conj1}, the MLDE solved by traces of higher powers of the BPS monodromy operator has the same order and Wronskian index as the one solved by the Schur index. 

In this paper, we have proved the generalised Schur partition function $\cZ_G(q;\alpha)$ yields a family of solutions for a fixed $N$-dependent order and Wronskian index MLDE for $G=USp(2N)$. As we will discuss in Section \ref{subsec:generalised_x2}, we suspect that this is still true for general $G$. Combining this expectation with Conjecture \ref{conj1}, we have that if there are two SCFTs whose Schur indices can be obtained from the same generalised Schur partition function, then the corresponding $\cI_k(q)$ satisfy MLDEs with the same order and Wronksian index.

Finally, we would like to comment that the precise relation between the generalised Schur partition function and traces of powers of the BPS monodromy matrix may include insertions of line defects. Such insertions are known to produce characters of twisted modules (as opposed to the vacuum/identity module) of the vertex operator algebra and may underlie the physical explanantion for the change of presentation that we discussed above. We do not explore this in the present work.

\subsection{Comments on the two-parameter generalisation}
\label{subsec:limit}

In \eref{eq:USp2N_genSchur_2par}, we proposed a two-parameter extension of the generalised Schur partiton function for the $USp(2N)$ SCFT with $2N+2$ fundamental hypermultiplets. This extension was based on the two-parameter contour integral \eref{JAp} and could be seen as a natural extension of the one-parameter $\cZ_{USp(2N)}(q;\alpha)$ written as an integral over $USp(2N)$. For this to have a consistent interpretation as a partition function of the associated 4d $\cN=2$ SCFT, we should be able to express this as a limit of the superconformal index. This was indeed the case for the one-parameter generalised Schur partition function which was obtained as a double scaling limit of the full superconformal index.

However, such a limit from the full superconformal index $\cI(p,q,t)$ to $\cZ_{USp(2N)}(q;\alpha,\beta)$ is not clear to us. In fact, a closer examination of \eref{eq:USp2N_genSchur_2par} reveals that realising $\cZ_{USp(2N)}(q;\alpha,\beta)$ from a specialisation of the variables $p,q,t$ alone is unlikely. This can be best understood by noticing that different roots of $USp(2N)$ are raised differently by the parameters $\alpha$ and $\beta$. In particular, the contribution of the roots $\pm 2e_i$ from the vector multiplets is raised to $\alpha$, while the contribution from the roots $\pm e_i\pm e_j$ is raised to $\beta$. At the level of the single letter indices, this would scale these roots differently in the adjoint character, making such a specialisation purely in terms of $p,q,t$ unlikely.

Recall that the $\alpha=0$ case of $\cZ_{USp}(2N)(q;\alpha)$ corresponded to the Schur index after giving all hypermultiplets a mass and turning on all Coulomb branch deformations. Analogously, setting $\beta=0$ in the two-parameter extension $\cZ_{USp(2N)}(q;\alpha,\beta)$ we obtain
\begin{equation}
    \cZ_{USp(2N)}(q;\alpha,0) \;\propto\; \prod_{i=1}^N \, \eta^2\oint \frac{{\rm d}z_i}{2\pi iz_i} \left(\eta^6\,\frac{\vartheta_1^2(2u_i)}{\vartheta_4^8(u_i)} \right)^\alpha ~.
\end{equation}
At the level of $USp(2N)$ representation theory, this can be seen as the restriction to the $SU(2)^{\otimes N} \subset USp(2N)$ root system. Furthermore, the structure of \eref{eq:USp2N_genSchur_2par} ensures that in the $\beta=0$ limit, we obtain the one-parameter generalised Schur partition function of a product of $N$ copies of $SU(2)$ SCFT with $4$ fundamental hypermultiplets.

We are not aware of any such deformation of the $USp(2N)$ SCFT and thus do not understand the physical origin, if any, of this statement. Therefore, we leave the proper four-dimensional interpretation of the two-parameter generalisation for future consideration.

\subsection{Generalisations in 4d and 2d}
\label{subsec:generalised_x2}

In this concluding section, we briefly formulate our expectations for the generalisations of the results we have presented in this work.

A natural question is whether this analysis can be extended to the generalised Schur partition functions of other 4d $\cN=2$ gauge theories. The empirical evidence indicates that the relation between $\cZ_G(q;\alpha)$ and the modular linear differential equation is structural and holds for arbitrary $G$ and corresponding matter content. Therefore, it is reasonable to posit that, for a fixed $G$ and matter content, the generalised Schur partition function $\cZ_G(q,\alpha)$ satisfies an MLDE of definite order and Wronskian index, thereby extending Proposition \ref{prop:USp2N}.

The main obstacle to extending our proof to the general case is technical. The key analytical bridge between $\cZ_{USp(2N)}(q;\alpha)$ and the MLDE was the contour integral representation of VVMFs that solve an MLDE with vanishing Wronskian index. However, an analogous series of elliptic transformations on the generalised Schur partition functions of other SCFTs does not yield contour integrals of the form \eref{JAp}. This is unsurprising, as \eref{JAp} themselves do not furnish the most general solution of the corresponding $\ell=0$ MLDE. Moreover, it is known that 4d Schur indices do not always satisfy MLDEs with vanishing Wronskian index. Furthermore, in many cases, the Schur index itself has a half-integer $q$-series expansion, and thus the corresponding MLDE is modular only under the congruence subgroup $\Gamma^0(2)$.

Thus, a uniform proof expressing $\cZ_G(q;\alpha)$ as a contour integral representation of an MLDE solution will likely require more general contour integral representations of MLDE solutions that account for these additional complications. We hope to address this problem in the future.

\acknowledgments{
PS would like to thank Sebastiano Garavaglia, Shlomo Razamat, and Sara Pasquetti for discussions related to this work.

RC acknowledges the support from the Department of Atomic Energy, Government of India, under project no. RTI4001. SM is supported by the Raja Ramanna Chair of the Department of Atomic Energy, Government of India. He would also like to acknowledge the warm hospitality and generous support of the International Centre for Theoretical Sciences (ICTS), where an early version of this project was initiated. PS is partially supported by the INFN and by the MIUR-PRIN grant 2022NY2MXY (finanziato dall’Unione europea - Next Generation EU, Missione 4 Componente 1 CUP H53D23001080006).
}

\appendix 

\section{Special functions}\label{app:special_functions}

In this appendix, we collect all the definitions, conventions, and various identities of the modular and elliptic functions that were used in the paper. We always take $q = e^{2\pi i \tau}$, with $\tau \in \mathbb{H}$. The $q$-Pochhammer symbol is defined as
\begin{align}
(a;q)_{n} = \prod_{k=0}^{n-1}(1-aq^k), \quad\quad (a;q)_{\infty} = \prod_{k=0}^{\infty}(1-aq^k), \quad |q| <1
\end{align}
For the infinite $q$-Pochhammer, $(a;q)_{\infty}$ we will drop the subscript and write it simply as $(a;q)$. We will often make use of the convention that if a $\pm$ sign appears in the infinite products such as $q$-Pochhammer symbols, we will take the product of all the $\pm$ signs. For example,
\begin{align}
(a^{\pm 1};q) = (a;q)(a^{-1};q)
\end{align}
The Dedekind $\eta$ function is defined as
\begin{align}
    \eta(\tau)  = q^{\frac{1}{24}}(q;q)
\end{align}
which is a modular form of weight $\frac{1}{2}$. Let us also recall that the holomorphic Eisentein series are modular forms of even weight $k\geq 2$ for the group $SL(2,\mathbb{Z})$ whose series expansions are given as
\begin{align}
    E_{2k}(\tau) = 1-\frac{4k}{B_k}\sum_{n=1}^{\infty}\frac{n^{2k-1}q^n}{1-q^n}
\end{align}
where $B_k$ are the Bernouli numbers. For $k=1$, the second Eisenstein series $E_2(\tau)$ is quasi-modular. It is used to define the Ramanujan-Serre derivative, which is a covariant derivative that acts on modular forms. On forms of weight $2k$, it is 
\begin{align}
    \cD = \frac{1}{2\pi i}\frac{\partial}{\partial\tau} - \frac{k}{6} E_2(\tau)
\end{align}

\subsection{Jacobi theta functions}
The four theta functions $\vartheta_i(u,\tau)$ are functions of two variables, $u \in \mathbb{C}$ and the modular parameter $\tau$. We will use the following definitions, where, as usual, $z = e^{2\pi i u}$ and $q = e^{2\pi i \tau}$
\begin{align}
        \vartheta_1(u,\tau) & = \sum_{n\in \mathbb{Z}}(-1)^{n-\frac{1}{2}}q^{\frac{1}{2}(n+\frac{1}{2})^2}z^{n+\frac{1}{2}}  = 2q^{\frac{1}{8}}\sum_{n=0}^{\infty}(-1)^nq^{\frac{n(n+1)}{2}}\sin{(2n+1)\pi u}\\[4pt]
        \vartheta_2(u,\tau) & = \sum_{n\in \mathbb{Z}}q^{\frac{1}{2}(n+\frac{1}{2})^2}z^{n+\frac{1}{2}} = 2q^{\frac{1}{8}}\sum_{n=0}^{\infty}(-1)^nq^{\frac{n(n+1)}{2}}\cos{(2n+1)\pi u}\\[4pt]
        \vartheta_3(u,\tau) & = \sum_{n\in \mathbb{Z}}q^{\frac{n^2}{2}}z^n  = 1+2\sum_{n=1}^{\infty}q^{\frac{n^2}{2}}\cos{2n\pi u}\\[4pt]
        \vartheta_4(u,\tau) &= \sum_{n\in \mathbb{Z}}(-1)^nq^{\frac{n^2}{2}}z^n= 1+2\sum_{n=1}^{\infty}(-1)^nq^{\frac{n^2}{2}}\cos{2n\pi u}
\end{align}
We can easily verify the periodicity properties of the theta functions from the expansions above. Apart from these series expansions, the theta functions also admit infinite product representations. Written compactly using $q$-Pochhammer symbols, we have the following infinite product expressions
\begin{align}
    \vartheta_1(u,\tau)&= \frac{q^{\frac{1}{8}}}{2\sin{\pi u}}(q;q)(z^{\pm 1};q)\\[4pt]
    \vartheta_2(u,\tau)&= \frac{q^{\frac{1}{8}}}{2\cos{\pi u}}(q;q)(-z^{\pm 1};q)\\[4pt]
    \vartheta_3(u,\tau)&= (q;q)(-q^{\frac{1}{2}}z^{\pm 1};q)\\[4pt]
    \vartheta_4(u,\tau)&= (q;q)(q^{\frac{1}{2}}z^{\pm 1};q)
\end{align}
where, for example, $(z^\pm;q)$ means the product $(z;q)(z^{-1};q)$ and:
\be
(z;q) \equiv (z;q)_\infty := \prod_{n=0}^\infty (1-zq^n)
\ee

Theta ``constants'' are defined as $\vartheta_i(\tau) := \vartheta_i(0,\tau)$. To make the notation less cumbersome, we will often supress the modular parameter and write the theta functions as $\vartheta_i(u) \equiv \vartheta_i(u,\tau)$ and the theta constants as $\vartheta_i \equiv \vartheta_i(\tau)$. 

Since $\vartheta_1(0,\tau)$ is identically zero, we have three theta constants which satisfy the relation
\begin{align}
\label{theta234}
    \vartheta_3^4(\tau) = \vartheta_2^4(\tau)+\vartheta_4^4(\tau)
\end{align}
The product of theta constants is given in terms of the Dedekind eta function as 
\begin{align}
\label{theta.eta}
    \vartheta_2(\tau)\vartheta_3(\tau)\vartheta_4(\tau) = 2\eta(\tau)^3
\end{align}
There exist several identities that involve products of the theta functions of different variables. For example, the following addition identity for $\vartheta_1$ will be useful
\begin{align}
\label{theta1.addition.identity}
    \vartheta_1(u_1+u_2,\tau)\vartheta_1(u_1-u_2,\tau) = \frac{\vartheta_1^2(u_1,\tau)\vartheta_4^2(u_2,\tau)-\vartheta_4^2(u_1,\tau)\vartheta_1^2(u_2,\tau)}{\vartheta_4(\tau)^2}
\end{align}
We will also make use of the so-called duplication identities for theta functions. For the case of $\vartheta_1$, we have the duplication identity
\begin{align}
\label{theta1.duplication}
    \vartheta_1(2u,\tau) = 2\frac{\vartheta_1(u,\tau)\vartheta_2(u,\tau)\vartheta_3(u,\tau)\vartheta_4(u,\tau)}{\vartheta_2(\tau)\vartheta_3(\tau)\vartheta_4(\tau)}
\end{align}
Similar expressions of addition and duplication identities for the other three theta functions are also well known and can be found, for example, in \cite{Lawden1989EllipticFA}. 

The theta constants are modular forms of weight $\frac{1}{2}$ under the congurence subgroup $\Gamma(2) \subset SL(2,\mathbb{Z})$. The modular lambda function is defined as the following ratio of theta constants, and has a series expansion in terms of half-integer powers of $q$
\begin{align}
    \lambda(\tau) = \frac{\vartheta_2^4(\tau)}{\vartheta_3^4(\tau)} = 16q^{\frac{1}{2}}\left(1-8q^{\frac{1}{2}}+44q+\cdots\right)
\end{align}
It is thus invariant under the action of $\Gamma(2)$. From the relation \eref{theta234}, we see that
\begin{align}
    1-\lambda(\tau) = \frac{\vartheta_4^4(\tau)}{\vartheta_3^4(\tau)}
\end{align}
Under the action of the full modular group, we have the transformations
\begin{align}
    \lambda(\tau+1) &= \frac{\lambda(\tau)}{1-\lambda(\tau)}\\
    \lambda\left(-\frac{1}{\tau}\right)&= 1-\lambda(\tau)
\end{align}

\subsection{Jacobi elliptic functions}
\label{subsec:Jacobi_elliptic}

Here we introduce our notations for the Jacobi elliptic functions $\text{sn},\text{cn},\text{dn}$. They can be defined in several equivalent ways, we will now define them in terms of ratios of theta functions. Let the elliptic modulus and the complementary modulus be defined using (where we suppress the explicit $\tau$ dependence on $\tau$ in both $k$ and $\lambda$) 
\begin{align}\label{eq:elliptic_modulus}
    k^2 = \lambda =  \frac{\vartheta_2^4(\tau)}{\vartheta_3^4(\tau)}, \qquad k'^2 = 1-k^2 =  \frac{\vartheta_4^4(\tau)}{\vartheta_3^4(\tau)}
\end{align}
These can be inversed to find the modular parameter $\tau$, which can be written in terms of the elliptic modulus as
\begin{align}
    \tau = i \frac{K'(k)}{K(k)}
\end{align}
where $K(k)$ and $K'(k)$ are complete elliptic integrals of the first kind
\begin{align}
    K(k) = \int_0^{\pi/2} \frac{d\theta}{\sqrt{1-k^2\sin^2{\theta}}}, \qquad K'(k) = K(k') 
\end{align}
the complete elliptic integral can be expressed in terms of ${}_2F_1$ as 
\begin{align}
    K(k) &= {}_2F_1\left(\frac{1}{2},\frac{1}{2},1, k^2\right) =  \frac{\pi}{2}\vartheta_3^2(\tau)
\end{align}
We are now ready to define the Jacobi elliptic functions. With
\begin{align}
    v = 2Ku = \pi\vartheta_3^2(\tau)u
\end{align}
they are given as the following ratios of theta functions
\begin{align}\label{eq:Jacobi_elliptic_def}
\text{sn}(v,k) &= \frac{\vartheta_3(\tau)}{\vartheta_2(\tau)}\frac{\vartheta_1(u,\tau)}{\vartheta_4(u,\tau)}\\[4pt]
\text{cn}(v,k) &= \frac{\vartheta_4(\tau)}{\vartheta_2(\tau)}\frac{\vartheta_2(u,\tau)}{\vartheta_4(u,\tau)}\\[4pt]
\text{dn}(v,k) &= \frac{\vartheta_4(\tau)}{\vartheta_3(\tau)}\frac{\vartheta_3(u,\tau)}{\vartheta_4(u,\tau)}
\end{align}

An equivalent way to define the Jacobi elliptic functions is directly via the (incomplete) elliptic integral of the first kind
\begin{align}
    F(\varphi,k) = \int_0^{\varphi} \frac{d\theta}{\sqrt{1-k^2\sin^2{\theta}}}
\end{align}
For $v = F(\varphi,k)$, the amplitude function is defined as $\text{am}(v,k) \equiv\varphi$. Then, we have the simple relations
\begin{align}
    \text{sn}(v,k) &= \sin{\text{am}(v,k)} = \sin\varphi\\[4pt]
    \text{cn}(v,k) &= \cos{\text{am}(v,k)} = \cos{\varphi}\\[4pt]
    \text{dn}(v,k) &= \frac{d }{dv}\text{am}(v,k) 
\end{align}
The Jacobi elliptic functions satisfy the elementary identities $\text{sn}^2(v,k)+\text{cn}^2(v,k)=1$ and $\text{dn}^2(v,k)+k^2\text{sn}^2(v,k)=1$. The elliptic functions are doubly periodic functions with the two periods. For the Jacobi sine, these are given as
\begin{align}
    \text{sn}(v+4K,k) = \text{sn}(v+2iK',k) = \text{sn}(v,k)
\end{align}
From the definition, we see that $\text{sn}(0,k)=0$. Some other useful values are
\begin{align}
    \text{sn}(K,k) = 1,\quad \text{sn}(2K,k)=0, \quad \text{sn}(iK',k) = \infty, \quad \text{sn}(K+iK',k) = k^{-1}
\end{align}
Finally, the derivatives of Jacobi elliptic functions can be obtained using identites relating the derivatives of theta functions. In particular, for the Jacobi sine we have
\begin{align}
\label{sn.derivative}
    \frac{d}{dv}\text{sn}(v,k) = \text{cn}(v,k)\,\text{dn}(v,k)
\end{align}

\section{General solutions of the second order $\ell=0$ MLDE}
\label{app:general_solution_MMS}
In this appendix we recall \cite{Mathur:1988gt,Naculich:1988xv} the most general solutions of the MMS equation. Though the MLDE is a differential equation in the modular parameter, it is easier to express the general solutions for the characters as functions of either the modular $j$ invariant of the modular $\lambda$ function. The MMS equation is
\begin{align}
    (\cD^2 +\mu E_4)\chi =0
\end{align}
Let us first perform the change of variables from $\tau$ to $K(\tau) := 12^3/ j(\tau)$. Denoting the differential operator $\theta_K = K\partial_k $, in the new variables we have the following hypergeometric ODE:
\begin{align}
    \left(\theta_K^2 -\frac{2K+1}{6(1-K)}\theta_K +\frac{\mu}{1-K}\right)\chi = 0
\end{align}
Noting that the free parameter $\mu$ is expressed in terms of $\alpha$ as
\begin{align}
\label{mu.alpha}
    \mu  = -\frac{(6\alpha+1)(6\alpha-1)}{144}
\end{align}
the two independent solutions for generic values of $\alpha$ are given as
\begin{align}
    \chi_0 = K^{\frac{1}{12}+\frac{\alpha}{2}}{}_2F_{1}\left(\frac{1}{12}+\frac{\alpha}{2},\frac{5}{12}+\frac{\alpha}{2},1+\alpha,K \right) \\[4pt]
    \chi_1 = K^{\frac{1}{12}-\frac{\alpha}{2}}{}_2F_{1}\left(\frac{1}{12}-\frac{\alpha}{2},\frac{5}{12}-\frac{\alpha}{2},1-\alpha,K \right) 
\end{align}
Whenever we have $\alpha \in \mathbb{Z}$, the two independent exponents differ by an integer and we have a resonance. In this case, we have logarithmic solutions. The two independent solutions can also be directly expressed in terms of the Meijer $G$-function as follows:
\begin{align}
    \chi_0  = G^{{1,1}}_{{2,2}}\left(K \,\,\Bigg|\,\, \begin{matrix}
1,\, \frac{2}{3} \\
\frac{1}{12}+\frac{\alpha}{2},\,\frac{1}{12}-\frac{\alpha}{2}
\end{matrix} \,\right) \\[4pt]
    \chi_1  = G^{{2,0}}_{{2,2}}\left(K \,\,\Bigg|\,\, \begin{matrix}
1,\, \frac{2}{3} \\
\frac{1}{12}+\frac{\alpha}{2},\,\frac{1}{12}-\frac{\alpha}{2}
\end{matrix} \,\right) 
\end{align}
These solutions are valid for all values of $\alpha$, including integers, in which case we find that the second solution above exhibits logarithmic behaviour.

In the main text, we have extensively used the modular $\lambda$ function. Keeping this in mind, let us also write down the general solutions in this variable. The MMS equation is now
\begin{align}
    \left(\partial_\lambda^2- \frac{2}{3}\frac{2\lambda-1}{\lambda(1-\lambda)}\partial_{\lambda} + \frac{4\mu(\lambda^2-\lambda+1)}{\lambda^2(1-\lambda)^2} \right)\chi = 0
\end{align}
whose two independent solutions are given, again relating $\mu$ to $\alpha$ using \eref{mu.alpha} as
\begin{align}
    \chi_0 = (\lambda(1-\lambda))^{\frac{1}{6}+\alpha}{}_2F_{1}\left( \frac{1}{2}+\alpha,\frac{1}{2}+3\alpha,1+2\alpha, \lambda \right)\\[4pt]
    \chi_1 = (\lambda(1-\lambda))^{\frac{1}{6}-\alpha}{}_2F_{1}\left( \frac{1}{2}-\alpha,\frac{1}{2}-3\alpha,1-2\alpha, \lambda \right)
\end{align}

The general solutions expressed in terms of Meijer $G$ functions, which are also valid for the logarithmic cases are
\begin{align}
    \chi = \lambda^{\frac{1}{6}-\alpha}(1-\lambda)^{\frac{1}{6}+\alpha} \left( c_1 G^{{1,1}}_{{2,2}}\left(\lambda  \,\,\Bigg|\,\, \begin{matrix}
\frac{1}{2}-\alpha,\, \frac{1}{2}+\alpha  \\
0,\,\ 2\alpha
\end{matrix} \,\right)  + c_2 G^{{2,0}}_{{2,2}}\left(\lambda  \,\,\Bigg|\,\, \begin{matrix}
\frac{1}{2}-\alpha,\, \frac{1}{2}+\alpha  \\
0,\,\ 2\alpha
\end{matrix} \,\right) \right)
\end{align}

\bibliography{Neq2andCFT}
\bibliographystyle{JHEP}

\end{document}